\documentclass{IEEEtran}
\ifCLASSINFOpdf

\else

\fi

\usepackage{amsmath,amsfonts,amssymb}
\usepackage{array}
\usepackage{textcomp}
\usepackage{stfloats}
\usepackage{url}
\usepackage{verbatim}
\usepackage{graphicx}
\usepackage{epstopdf}
\usepackage{cite}

\usepackage{subfigure}
\usepackage{array}
\usepackage{epstopdf}
\usepackage{mathrsfs}
\usepackage{enumerate}
\usepackage{bm}
\usepackage{dsfont}
\usepackage{pifont}
\usepackage{setspace}
\usepackage{multirow}
\usepackage{color}

\usepackage{threeparttable}
\usepackage{adjustbox}
\usepackage{tabularx}
\usepackage{booktabs}
\usepackage{makecell}

\usepackage{algorithm}
\usepackage{algorithmicx}
\usepackage[noend]{algpseudocode}

\algnewcommand{\LineComment}[1]{\State {\color{blue}\(\triangleright\) #1}}
\usepackage[linkcolor=blue,citecolor=blue,colorlinks=true,urlcolor=blue]{hyperref}

\newcommand{\x}{{\boldsymbol x}}
\newcommand{\y}{{\boldsymbol y}}
\newcommand{\bmhatx}{{\boldsymbol{\hat{x}}}}
\newcommand{\bmhaty}{{\boldsymbol{\hat{y}}}}
\newcommand{\bmcheckx}{{\boldsymbol{\check{x}}}}
\newcommand{\bmchecky}{{\boldsymbol{\check{y}}}}

\begin{document}

\title{Loss-Resilient Semantic Communication over Packet-Loss Networks at Extreme-Low Bandwidth}

\author{Shengshi Yao,~\IEEEmembership{Member,~IEEE},
        Jincheng Dai,~\IEEEmembership{Member,~IEEE},
        Sixian Wang,~\IEEEmembership{Member,~IEEE},\\
		Guo Lu,~\IEEEmembership{Member,~IEEE},
		Kai Niu,~\IEEEmembership{Member,~IEEE},
		Wenjun Xu,~\IEEEmembership{Senior Member,~IEEE},\\
        Wenjun Zhang,~\IEEEmembership{Fellow,~IEEE},
        and Ping Zhang,~\IEEEmembership{Fellow,~IEEE}

\thanks{This work was supported in part by the National Key Research and Development Program of China under Grant 2024YFF0509700, in part by the National Natural Science Foundation of China under Grant 62371063, Grant 62471290, Grant 62501075, and Grant 92467301, in part by the Beijing Natural Science Foundation under Grant L232047, in part by Postdoctoral Fellowship Program of CPSF under Grant Number GZB20250810, in part by China Postdoctoral Science Foundation under Grant Number 2025M783515, and in part by the Beijing Nova Program. (\textit{Corresponding author: Jincheng Dai.})}

\thanks{Shengshi Yao, Jincheng Dai, Kai Niu, Wenjun Xu and Ping Zhang are with Beijing University of Posts and Telecommunications, Beijing 100876, China (e-mail: daijincheng@bupt.edu.cn).}

\thanks{Sixian Wang, Guo Lu and Wenjun Zhang are with Shanghai Jiao Tong University, Shanghai 200240, China.}

\thanks{Open-source code and data are available on-line at: \url{https://github.com/semcomm/ResiGLC}.}

}

\maketitle

\begin{abstract}

In extreme-low bandwidth network scenarios, generative semantic codecs have emerged as promising solutions to reduce bandwidth cost for visual communications. However, these learned codecs are usually optimized solely for compression efficiency and thus not robust against transmission errors. Corruptions due to packet-loss among these highly compact generative latent representations often cause more critical degradation in fidelity and realism, intensified by the severe error propagation across the latent contexts and multi-step decoding process. In this paper, we propose ``\emph{ResiGLC}'', a novel loss-resilient generative latent coding framework designed for robust semantic communication over extreme-low bandwidth packet-loss networks. Motivated by the inherent goal-consistency between generation and compression, we sufficiently exploit the impressive in-context predictive capabilities of language models. Integrated with the masked learning strategy, our model supports arbitrary context modeling of latent codes, which could mitigate the error propagation and handle unpredictable packet loss patterns. At the receiver, a progressive resilient decoding pipeline is presented, which leverages both the contextual relationship of the latent codes and the multi-modal semantic prior in the generative latent space, separately. By jointly optimizing toward both compression efficiency and packet-loss resilience, our proposed progressive decoding mechanism offers graceful performance when dealing with dynamic packet losses. Through extensive experimental evaluations, we establish that under packet-loss network conditions, \emph{ResiGLC} can effectively improve the loss-resilience in terms of perceptual fidelity and realism qualities with extreme-low bandwidth cost. 
\end{abstract}

\IEEEpeerreviewmaketitle

\section{Introduction}\label{section_introduction}

\IEEEPARstart{R}{obust} {visual communication remains a critical challenge, as conventional systems struggle to deliver a consistently plausible Quality of Experience (QoE) under severe network conditions. The problem is particularly pronounced in scenarios such as deep-sea, deep-space telemetry, and emergency communications, where the available wireless network bandwidth is extremely limited and the network conditions are highly dynamic and volatile~\cite{wu2000transporting}. This necessitates a paradigm shift toward one-shot content delivery frameworks.}

Taking image communication as a case study, conventional image codecs optimized toward minimizing pixel-wise distortion~\cite{bpg, balle2018variational, cheng2020learned} typically operate in the mid-to-high bitrate regime ($>0.15$ bits per pixel (bpp)).
These codecs tend to produce blurring and artifacts as the bitrate decreases.
Therefore, in extreme-low to low bandwidth scenarios (typically $<0.15$ bpp), generative compression has been recognized as the general approach to reduce the bandwidth consumption while also ensuring high fidelity.
Motivated by Generative Adversarial Networks (GAN)~\cite{goodfellow2020generative} and diffusion models~\cite{ddpm}, generative image codecs~\cite{mentzer2020high,msillm,diffeic,careil2023towards}, often referred to as \emph{perceptual image codecs}, learn to enhance the perceptual quality with significantly fewer bits.
They achieve better perceptual quality and rate efficiency at the cost of higher distortion. 
With the objective of high-{\textit{realism}} and high-{\textit{fidelity}} at low bitrate, generative latent coding~(GLC) has been emerging as a dominant paradigm that performs transform coding at either quantized~\cite{jia2024generative} or continuous~\cite{careil2023towards} generative latent space to align with human perception. 
This allows noticeable bitrate savings by leveraging the latent prior.
These promising works have catalyzed the development of image semantic communications over extreme bandwidth-limited channels.
{While the employment of generative image codecs inevitably introduces evident computational complexity at the decoder, this paradigm fits these applications, which are fundamentally featured by asymmetric computational resources. Specifically, the transmitters are often severely resource-constrained devices, such as autonomous underwater vehicles, whereas the receivers are typically ground stations equipped with abundant computational resources.}

However, existing generative image codecs are typically developed under the assumption of error-free network transmission. 
To mitigate potential transmission errors, these codecs typically resort to application- or transport-layer error control mechanisms. Among them, Automatic Repeat reQuest (ARQ) and Forward Error Correction (FEC) coding~\cite{nafaa2008forward,dred,rlafec} are widely used. 
On one hand, ARQ serves as a reactive mechanism initiated by the receiver, by sending a feedback signal for retransmission. 
Nevertheless, in the aforementioned extreme environments, the inherently long round-trip times (RTTs) render retransmission latency non-deterministic and prohibitive. 
On the other hand, the effect of FEC, as a proactive error correction mechanism, is correlated with the proportion of redundancy, in which case the sender needs to estimate the packet loss ratio in advance. Otherwise, it risks either allocating excessive redundancy or suffering decoding failures that still necessitate retransmission.
More critically, the transmission overhead of FEC becomes unaffordable under extreme-low bandwidth conditions.
Parallel to ensuring reliable transmission of compressed source data, Joint Source-Channel Coding (JSCC) has emerged as a distinct branch of research for robust transmission. Mapping source data directly to transmitted symbols via deep neural networks, deep JSCC~\cite{bourtsoulatze2019deep,esfahanizadeh2026block} shows prospects in improving end-to-end transmission performance. However, its deployment faces fundamental compatibility hurdles with existing standardized digital communication systems.

Therefore, there exists a pressing need for novel coding paradigms that offer inherent resilience within bandwidth-limited and volatile environments~\cite{he2024toward, huang2025d}. {Compared to the aforementioned mechanisms, this paper advances packet loss concealment (PLC), a passive error control mechanism, for generative image codecs, which strikes a balance between compression efficiency and loss-resilience. It thereby ensures a more deterministic latency by minimizing the reliance on latency-inducing retransmissions, and also circumvents excessive FEC redundancy in the event of unpredictable packet loss.}

\begin{figure}[t]
	\setlength{\abovecaptionskip}{0.cm}
	\setlength{\belowcaptionskip}{-0.cm}
	\centering
	\includegraphics[width=\columnwidth]{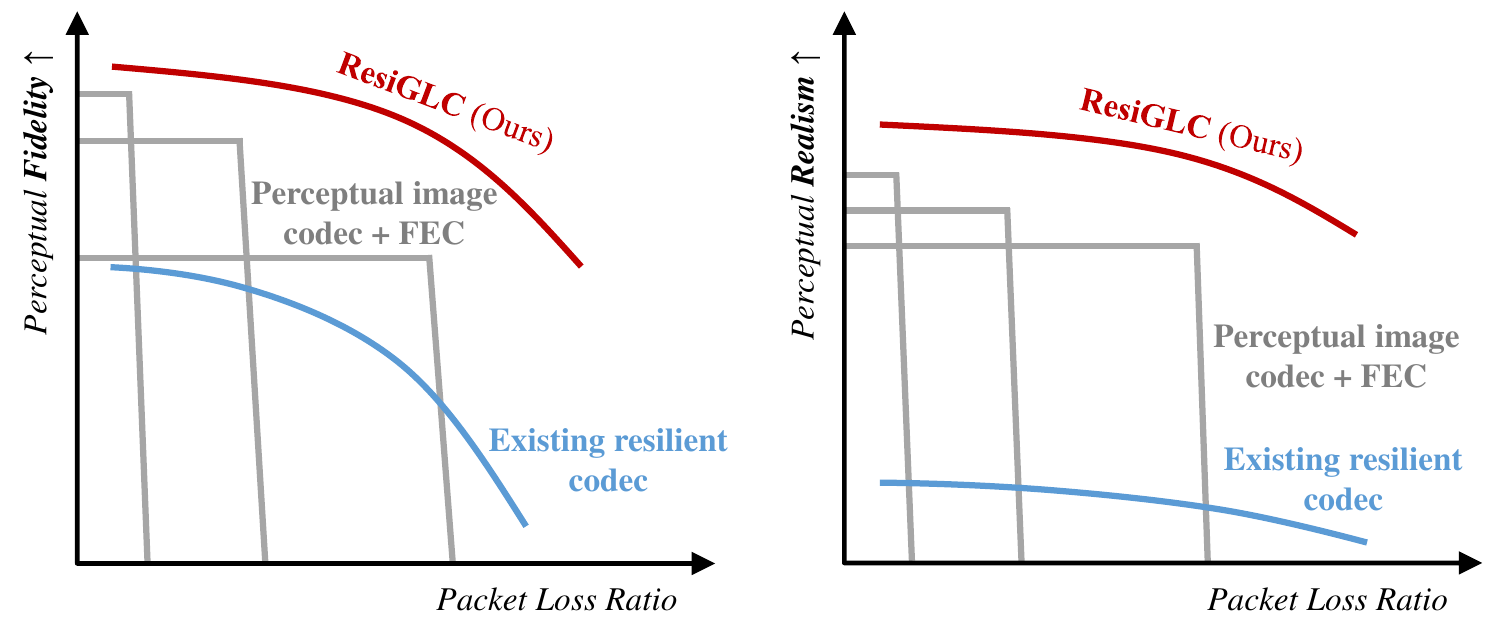}
	\caption{An example of loss resilience in terms of perceptual fidelity and realism under different loss ratios, where the upward arrow ``$\uparrow$'' indicates that higher value of the metric is favorable. 
		The figures are illustrative, and more results as well as the analysis can be found in Section~\ref{section_experiments}.
	}\label{fig_summary}
	\vspace{-1em}
\end{figure}

Specifically, the challenges for the generative image codecs to accomplish the error-resilience in bandwidth-limited scenarios are twofold: 
\begin{enumerate}
\item Extreme utilization of the contextual dependencies among latent codes may result in severe error propagation in the event of packet loss. The eventual image reconstruction quality is highly sensitive to packet loss, exhibiting a pronounced \emph{cliff effect} (illustrated by the gray curves in Fig.~\ref{fig_summary}). 

\item Perceptual codecs that employ a conditional generative model generally premise on access to complete and precise knowledge of the conditions. 
Transmission error leads to incomplete conditions, which undermines content consistency. More critically, it may hinder the generation process, preventing samples from converging to the learned manifold where realistic data resides.
\end{enumerate}

This paper proposes~\emph{ResiGLC}, a novel error-resilient generative latent coding framework. 
It operates as a multi-modal semantic transmission framework that conveys both compact image features and global textual semantics. 
To address the aforementioned challenges, rather than exhaustively exploiting contextual dependencies for extreme compression, we strategically preserve partial contextual redundancy at the sender. It mitigates error propagation while maintaining a favorable balance with compression efficiency. 
{Motivated by masked modeling strategies~\cite{maskgit,gao2025cross}}, a masked modeling Transformer (MMT) is designed to achieve \emph{arbitrary contextual modeling}. 
The multi-modal scheme further bolsters resilience, enabling robust reconstruction provided that any modality of condition signals is available. In worst cases where all latent codes are unrecoverable, the text description alone still yields semantically consistent and perceptually plausible reconstructions. 
In other cases, the fidelity will be further enhanced with more latent codes received. 
To prevent incomplete conditions from driving the generation off the diffusion prior, we introduce a \textit{progressive resilient decoding} mechanism at the receiver. 
Instead of directly exposing incomplete latent codes to the diffusion model, the receiver first leverages the learned contextual prior to predict the lost latent codes. Subsequently, the restored codes act as conditions to guide the token sequence generation in the generative latent space for image synthesis.
A corresponding multi-phase training strategy is devised to optimize this pipeline.

We verify the performance advantages of \emph{ResiGLC} under diverse packet-loss channel setups, including random packet loss traces as well as Markovian channel loss traces as the composite of realistic channels. 
Results demonstrate that the proposed \emph{ResiGLC} strikes a commendable balance between efficiency and resilience against packet loss. 
In contrast to FEC-based strategy, \emph{ResiGLC} does not presume any knowledge or estimation of packet loss at the sender. Its progressive resilient decoding ensures graceful degradation in terms of perceptual fidelity and realism, without the need for retransmission or extra redundancy. 
Existing resilient image codecs are mostly distortion-oriented, e.g., \cite{sha2025towards,wang2025resicomp}, which struggle to maintain perceptual quality at extreme-low bandwidth cost (the blue curves in Fig.~\ref{fig_summary}).
We highlight the attractive properties of our \emph{ResiGLC} compared to existing perceptual or resilient codecs in Fig.~\ref{fig_summary}. Please refer to Section~\ref{section_experiments} for more numerical results.

The remainder of this paper is organized as follows.
In the next section, we briefly review the related work on low-bitrate image codecs and classical error-resilience techniques.
Section~\ref{section_method} presents the framework design and stage-wise decoding pipeline of \emph{ResiGLC}.
Section~\ref{section_experiments} presents experiments on our proposed method to illustrate the performance gain, comparing with perceptual image codecs and existing resilient image codecs.
Finally, Section~\ref{section_conclusion} concludes the paper.

\textit{Notational Conventions:} {Thin letters (e.g., $x$) denote scalars, lowercase bold letters (e.g., $\x$) denote vectors, and uppercase bold letters (e.g., $\mathbf{X}$) denote matrices. Calligraphic letters (e.g., $\mathcal{X}$) are used to denote sets or loss functions. $p(\x)$ denotes the probability density function (PDF) of a random variable $\x$. $\mathcal{N}(\mu, \sigma^2)$ represents the Gaussian distribution with mean $\mu$ and variance $\sigma^2$. $\mathbb{E}[\cdot]$ denotes the mathematical expectation, and $\mathbb{N}_+$ represents the set of positive integers. Furthermore, the diacritics $\bmhatx$ and $\bmcheckx$ typically signify quantized/decoded variables and predicted variables, respectively.}

\section{Related Work}\label{section_related}

\subsection{Image Codecs toward Extreme-Low Bitrate}

Learned Image Compression (LIC) methods using neural networks have shown superior performance over traditional methods such as BPG~\cite{bpg}. One of the most trending LIC frameworks exploits variational auto-encoder~(VAE) to regularize the distribution of latent features, whose context is modeled by an entropy model~\cite{balle2018variational,cheng2020learned}. These codecs are optimized using a Lagrange objective as
\begin{equation}\label{eq_rd_obj}
  \mathcal{L} = \lambda D + R,
\end{equation}
where $D$ quantifies the distortion, e.g., the mean squared error of two images $\text{MSE}(\x, \bmhatx)$, $\lambda$ is the weighting factor. 
However, in the low-bitrate regime, the information loss is severe, and the reconstructions exhibit mode averaging behavior, degrading visual quality for humans. 

To mitigate this phenomenon, perceptual loss $D_P$, as a complement of~\eqref{eq_rd_obj}, is adopted to regularize feature-wise distance, e.g., LPIPS~\cite{lpips}. It results in a rate-distortion-perception~(\mbox{RDP}) function 
\begin{equation}\label{eq_rdp_obj}
  \mathcal{L} = \lambda_P D_P + \lambda D + R,
\end{equation}
which relaxes the constraint of low-level distortion.
The adversarial training using GAN~\cite{mentzer2020high,msillm} encourages the codecs to generate details alike to real images, enhancing perceptual \emph{realism}.
MS-ILLM~\cite{msillm} introduced a non-binary discriminator conditioned on local image representations, significantly improving the realism of generative compression. 
Some works employed more powerful diffusion models as generative models~\cite{careil2023towards, theis2022lossy, diffeic}.

The key challenge of these generative image codecs lies in the fidelity to the original image (reference-based distance), despite good realism which mathematically correlates to statistical feature distance. 
Codecs that generate images from scratch or from text~\cite{lei2023text} compromise the image fidelity despite extreme-low bitrate. 
Blau et al.~\cite{blau2018perception} explain this phenomenon by highlighting a fundamental trade-off between perceptual quality and distortion.
Technically, DiffEIC~\cite{diffeic} exploits the generative capability of pre-trained diffusion models and injects guidance to improve fidelity.
GLC~\cite{jia2024generative,careil2023towards} employs a human-perception aligned generative latent space of VAE or vector-quantized VAE~\cite{vqvae}, trading fidelity for better perceptual realism at extreme-low bitrates.

\subsection{Error-Resilience Techniques}

To transmit highly-compact image representations, various approaches have been developed to enhance the robustness against random data errors and erasures, including forward error correction (FEC), packet loss concealment (PLC) and joint source-channel coding (JSCC).

1) \textit{Forward error correction (FEC)} is a fundamental method used to protect the compressed bitstream from transmission errors. It works by adding redundancy to the data at the sender side, either at the application or transport layer. Specifically, FEC encodes $N_k$ data packets and adds $N_r$ parity packets, such that the original data packets can be recovered if any subset of $N_k (1 + r)$ packets out of the total $(N_k + N_r)$ packets are received. $r$ denotes the redundancy ratio required for correct data recovery. Common FEC codes include Reed-Solomon codes in storage systems, and fountain codes in multimedia broadcasting~\cite{nafaa2008forward}. 
However, the potentially inadequate redundancy in environments worse than expected causes transmission breakdown, while excessive redundancy compromises efficiency. 
Valin et al.~\cite{dred} investigated low bitrate redundancy coding in speech coding to reduce the FEC overhead.
Learning-based FEC strategies~\cite{rlafec} have been proposed to adaptively select redundancy level for different video frames based on channel conditions.  
These works partially mitigate the limitations of traditional FEC concerning both efficiency and effectiveness. 
However, FEC cannot guarantee perfect recovery, as its effectiveness breaks down under severe dynamic packet loss. 

2) \textit{Packet Loss Concealment (PLC)} is a long-lasting topic in media transmission. 
A common category of PLC is redundancy coding, which has been widely applied in speech coding~\cite{opus}. Different from the parity check in FEC, it encapsulates partial but key content in adjacent data packets to maintain a minimum quality of reconstruction. However, the concealment also fails if both the data and redundancy are lost.
In recent years, data-driven PLC methods using neural networks have been developed~\cite{sha2025towards,wang2025resicomp,yao2025soundspring,jiang2023latent,cheng2024grace} for resilient audio and image coding. 
PLC is either integrated within the codec or implemented after the decoding, exploiting the inherent correlation of data to predict the missing data (features).
Nevertheless, we find that the resilience struggles in balancing the fidelity and perceptual realism at low bandwidth cost. 

3) \textit{Joint Source-Channel Coding (JSCC)} is another active area of research toward resilient transmission, which directly maps source data to transmitted symbols. Existing studies have demonstrated significant robustness against lossy communication channels and superior end-to-end coding gain due to joint source-channel processing~\cite{bourtsoulatze2019deep,esfahanizadeh2026block}. Digital modulation methods on top of JSCC have been proposed to enhance the compatibility with current digital communication systems~\cite{huang2025d, zhang2025from}. 
However, the strong coupling behaviour between physical layer and application layer, and the protocol mismatch hinders the practical deployment of JSCC.

In summary, to fill the gap of resilient image coding at extreme-low bandwidths within current digital communication systems, exploring inherent resilience for a low-bitrate codec is a critical necessity to handle inevitable packet loss.

\begin{figure*}[ht]
	\setlength{\abovecaptionskip}{0.cm}
	\setlength{\belowcaptionskip}{-0.cm}
	\centering
	\includegraphics[width=2\columnwidth]{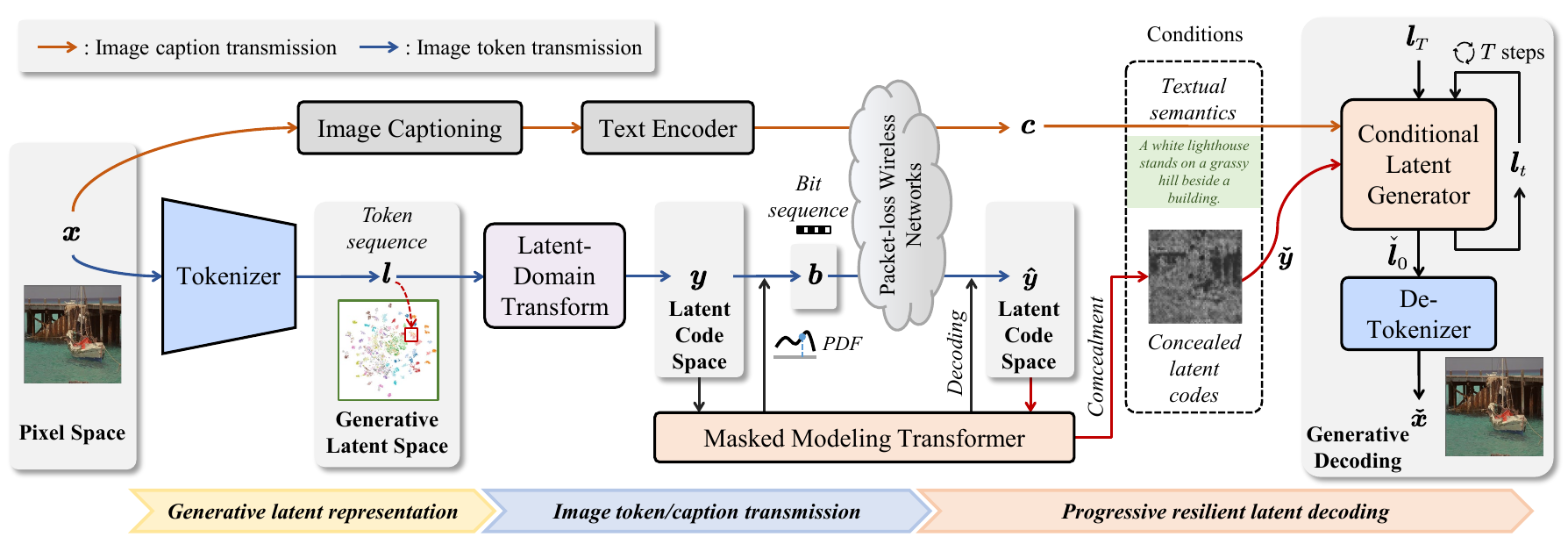}
	\caption{Overview of ResiGLC. 
	Two data streams are compressed and transmitted, including textual semantics and the latent codes extracted from the token sequence in generative latent space. 
	The Masked Modeling Transformer (MMT) models the contextual dependencies of latent codes, which are exploited for both entropy coding (produce probabilistic density function (PDF)) and loss concealment (predict lost codes) in decoding stage I.
	The conditional latent generator leverages the text-aided semantic latent prior and the concealed codes as condition signals, restoring the token sequence by multiple diffusion steps in decoding stage II.
	}\label{fig_overview}
\end{figure*}

\section{Method}\label{section_method}

This paper presents a novel error-resilient generative semantic communication framework that incorporates generative modeling of latent space in handling lossy transmission in bandwidth-limited scenarios.

\subsection{Overview}

We recall that the objective function~\eqref{eq_rd_obj} of an image codec, is employed to seek a rate-distortion tradeoff, where the rate of the compressed vectors $\bmhaty$ 
\begin{equation}
	R \triangleq R_{\bmhaty} = -\log_2p(\bmhaty),
\end{equation} 
is measured by an entropy model. We empirically find that it fails to yield high-quality image reconstructions at extreme-low bitrates, primarily due to excessive information loss of $\bmhaty$. Specifically, both \textit{fidelity} and \textit{realism} appear to degrade concurrently. This phenomenon manifests the inherent RDP trade-off in~\eqref{eq_rdp_obj}, which becomes particularly acute under severe bandwidth constraint.

Thus, following GLC, ResiGLC performs transform coding in the generative latent space.
The overview of the proposed framework is shown in Fig.~\ref{fig_overview}.
It begins with an image tokenizer $\mathcal E$, which transforms pixels $\x$ to the token sequence $\boldsymbol{l}$.
The tokens reside in the latent space which exhibits better scaling properties with respect to the spatial dimensionality. 
Then, latent codes $\y \in \mathbb R^{N \times C}$ are extracted from the token sequence via an analysis transform $g_a$, where $N, C$ denotes the spatial and the channel dimension. 
The sender is going to compress $\y$ into an extreme low-bitrate representation $\boldsymbol{b}$ to meet bandwidth requirements.  
Firstly, a scalar \textbf{quantizer} $Q$ quantizes them to the discrete form $\bmhaty$, rounding to the nearest integers element-wise. 
A \textbf{slice partition} module (omitted in the figure) divides $\bmhaty$ into $K$ code slices $\mathcal Y=\{\mathcal Y_1, \cdots, \mathcal Y_K\}$. 
Each code slice is separately entropy-encoded to a bitstream $\mathbf{b}_k, k=1,\cdots,K$, with contextual dependency determined by a context map $\mathbf{G}$. The contextual modeling is implemented by a \textbf{masked modeling Transformer} (MMT).
Finally, the coded bitstreams are packetized and then transmitted over a packet lossy wireless channel.

At the receiver, ResiGLC aims to reconstruct high-quality images from the received latent codes, which are corrupted versions of $\bmhaty$ in the context of packet-lossy transmission. 
Assuming the Open Systems Interconnection model, packet loss stems not only from severe channel fading at the physical layer~\cite{5341334} but also the buffer overflow at higher layers. 
The image decoder actually contends with an erasure channel, where erasures occur arbitrarily or in bursts. 
To model this, the transfer function of packet lossy channel $W(\mathcal Y, \kappa)$ is characterized by binary-valued loss traces, where the parameter set $\kappa$ depicts the trace distribution.

The receiver strives to reconstruct the image via a two-stage progressive decoding procedure.
In stage I, it firstly entropy decodes the latent codes from the received packets. Then, the latent code concealment reuses the MMT to predict the lost or undecodable codes. 
In stage II, a conditional latent generator restores the token sequence conditioned on the MMT-compensated latent codes as well as a global text description $\boldsymbol{c}$ of the image extracted via an image captioning module $f_{\text{caption}}$. Finally, the de-tokenizer maps the restored token sequence $\bm{\check{l}}_0$ from the generative latent space back to the pixel space, with $\bmcheckx = \mathcal{D}(\bm{\check{l}}_0)$.

In the following subsections, we will detail the workflow of the dual-functional MMT, and the progressive resilient decoding pipeline.

\subsection{Masked Modeling Transformer for Feature Compression and Concealment}

Current high-efficient neural image codecs obtain remarkable coding gain by designing powerful context models to better model $p(\hat \y)$. 
The most representative context mode is autoregressive modeling that decomposes the joint likelihood of latent codes into the product of a series of conditional likelihoods:
\begin{equation}
  p(\mathcal Y_1, \cdots, \mathcal Y_K) = \prod_{k=1}^{K}{p(\mathcal Y_k | \mathcal Y_1, \cdots, \mathcal Y_{k-1})},
\end{equation}
based on the information-theoretic fact that the conditional entropy $H(\mathcal Y_k | \mathcal Y_1, \cdots, \mathcal Y_{k-1})$ is not larger than the marginal entropy $H(\mathcal Y_k)$.
However, $\mathcal Y_k$ fails to be recovered by entropy decoding at the receiver if any of the contextual codes $\mathcal{Y}_{\text{ctx}} = \left\{\mathcal Y_1, \cdots, \mathcal Y_{k-1} \right\}$ is not received. 
Existing error resilient codecs have attempted to limit the utilization of context to trade off coding efficiency for resilience.

\begin{figure}[t]
	\setlength{\abovecaptionskip}{0.cm}
	\setlength{\belowcaptionskip}{-0.cm}
	\centering
	\includegraphics[width=\columnwidth]{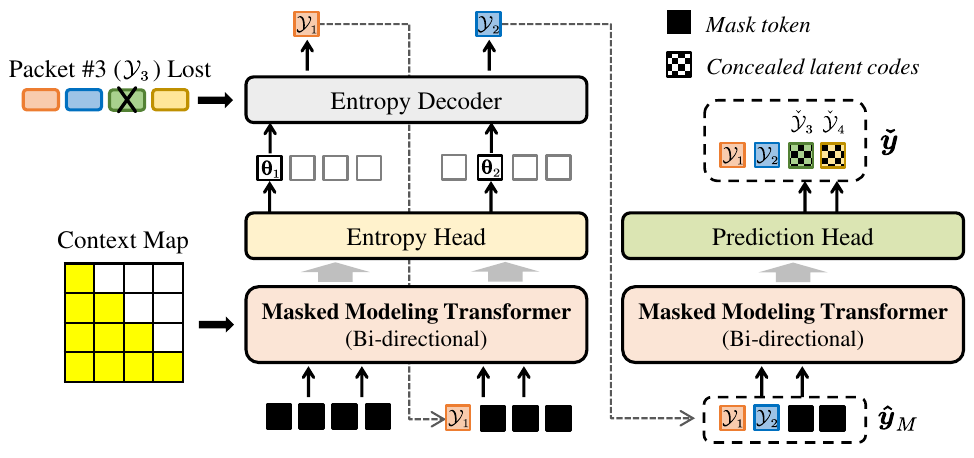}
	\caption{Latent code decoding and concealment via dual-functional masked modeling Transformer.
	}\label{fig_MMT}
\end{figure}

\begin{algorithm}[t]
	\caption{ResiGLC Sender}\label{algorithm_sender}
	\textbf{Input:} Input image $\bm{x}$, $\text{partitioner}$, text encoder, image caption module $f_{\text{caption}}$, packet number for latent codes $K \in \mathbb{N}_+$, mask token $\bm{m}_{\text{mask}}$, and context map $\mathbf{G} \in \{0,1 \}^{K \times K}$
	
	\textbf{Output:} Set of packets $\{\mathcal{P}_1, \cdots, \mathcal{P}_{K+1}\}$
	\begin{algorithmic}[1]
		\Function{ContextualModel}{$\mathcal{Y}, i, \mathbf{G}, \boldsymbol{r}$}
		\State $\y_{i, \text{ctx}} \gets \emph{ones}(N, C) \cdot \bm{m}_{\text{mask}}$
		\For {$j \in \{1, \cdots, i-1\}$} \Comment{Collect context}
		\State $\bm n_j \gets  \text{partitioner.get\_locations(step=j)}$ 
		\State $\y_{i, \text{ctx}}[\bm n_j, :] \gets \mathcal{Y}_j$ if $\mathbf{G}[i,j] \cdot \boldsymbol{r}[j] = 1$
		\EndFor
		\State  $\mathbf{C}_i \gets f_{\text{MMT}}\left( \y_{i, \text{ctx}} \right)$
		\State \Return context vector $\mathbf{C}_i$
		\EndFunction
		\Function{ResiGLC Encoder}{$\x, K, \bm{m}_{\text{mask}}, \mathbf{G}$}
		\LineComment{Image tokenization to generative latent space}
		\State $\bm l \gets \mathcal E(\x)$  
		\LineComment{Latent code extraction and quantization}
		\State $\bmhaty \gets Q\left(g_a(\bm l)\right)$  
		\State $\mathcal{Y} \gets \{\mathcal{Y}_1, \cdots, \mathcal{Y}_K\} \gets  \text{partitioner.get\_slices}(\bmhaty, K)$
		\For {$k \in \{1, \cdots, K\}$}
		\LineComment{Extract contexts of slice $k$}
		\State $\mathbf{C}_k \gets \Call{ContextualModel}{\mathcal{Y}, k, \mathbf{G}, \emph{ones}(K)}$
		\LineComment{Gaussian parameters estimation}
		\State $\bm{\mu}_k, \bm{\sigma}_k \gets h_{\text{entropy}}\left(\mathbf{C}_k[\bm n_k, :] \right)$
		\LineComment{Slice-based entropy encoding \& packetization}
		\State $\mathcal{P}_k \gets \text{entropy\_encoder}(\mathcal{Y}_k, \bm{\mu}_k, \bm{\sigma}_k)$
		\EndFor
		\LineComment{Encoding text in the last packet}
		\State $\mathcal{P}_{K+1} \gets \text{text\_encoder}(\boldsymbol{c}=f_{\text{caption}}(\x))$ 
		\State \Return Set of packets $\{\mathcal{P}_1, \cdots, \mathcal{P}_{K+1}\}$
		\EndFunction
	\end{algorithmic}
\end{algorithm}

Motivated by these prior works, a dual-functional masked modeling Transformer is employed to perform 1) entropy modeling for efficient coding at the transmitter end, and 2) loss concealment for resilient decoding at the receiver end. 
Specifically, the latent codes $\bmhaty$ are variationally modeled as multi-variate Gaussian variables, whose probabilistic density functions (PDF) are established in a factorized manner by
\begin{equation}
	p(\bmhaty) = \prod_{i=1}^N p(\hat{y}_i|\bmhaty_{i,\text{ctx}}) = \left(\mathcal{N}(\mu_i, \sigma_i^2)\right) \left(\hat{y}_i |\bmhaty_{i,\text{ctx}}\right),
\end{equation}
which are conditioned on the corresponding contextual latent codes $\bmhaty_{i,\text{ctx}}$. 
The detailed workflow of ResiGLC sender is illustrated in Algorithm~\ref{algorithm_sender}.
As described previously, the entropy modeling is implemented slice-wisely for acceleration, computing 
$\bm{\mu}_k = \left\{\mu_i\right\}_{i \in \bm n_k}$ and $\bm{\sigma}_k = \left\{\sigma_i\right\}_{i \in \bm n_k}$ for the $k$-th slice. $\bm n_k$ encapsulates the indices of latent codes within the $k$-th slice, which is determined by the slice partitioner.

At the receiver, ResiGLC firstly tries to entropy decode $\bmhaty$ using the exact contextual relationship at the sender. Secondly, for the lost latent codes, the MMT is also employed to predict the original latent codes from the received ones.
Assuming $\mathbf{M} \in \{0, 1\}^{N \times C}$ is a binary-valued matrix indicating the state of the latent code in every spatial location whether correctly decoded or not, the lost latent codes are replaced by a learnable mask embedding $\bm{m}_{\text{mask}} \in \mathbb{R}^{C}$.
The MMT accepts $\bmhaty_M = \bmhaty \cdot (\bm{1} - \mathbf{M}) + \bm{m}_{\text{mask}} \cdot \mathbf{M}$ for code prediction.
Thus, the resulting concealed latent codes can be formulated as
\begin{equation}
	\check{\bm{y}} = h_{\text{PLC}}\left(f_{\text{MMT}}(\bmhaty_M)\right) \cdot \mathbf{M} + \bmhaty_M \cdot (\bm{1} - \mathbf{M}),
\end{equation}
which are later employed in stage II to enrich the content consistency. The workflow of stage I resilient decoding is summarized in Fig.~\ref{fig_MMT}.
The Transformer layers are reused for both tasks, after which an entropy head $h_{\text{entropy}}$ predicts the distribution parameters $\boldsymbol{\theta}_k = [\bm{\mu}_k, \bm{\sigma}_k], k=1,\cdots,K$, and a prediction head $h_{\text{PLC}}$ predicts the latent codes to fill the masked positions.

\subsection{Spatial-aware Conditional Latent Generation}

Despite maintaining good spatial consistency within the latent codes $\y$, the extremely compressed representation still leads to significant information loss.
Therefore, instead of directly decoding the latent codes, ResiGLC tries to incorporate the prior knowledge of the token sequence in generative latent space, with the supervision of latent codes.
As shown in the right of Fig.~\ref{fig_overview}, apart from the concealed latent codes $\bmchecky$, we also leverage the global text description $\boldsymbol{c}$ of the image extracted at the encoder as an extra condition signal.
Thus, the conditional latent generation could effectively take advantage of the diffusion prior as well as the spatial details.

\begin{algorithm}[t]
	\caption{ResiGLC Receiver}\label{algorithm_receiver}
	\textbf{Input:} Packet set $\mathcal{P}$, number of packets $K$ for latent codes, mask token $\bm{m}_{\text{mask}}$, context map $\mathbf{G}$, packet reception indicator~$\boldsymbol{r} \in \{0,1\}^{K+1}$, number of diffusion steps $T$, and text guidance strength $w$.
	
	\textbf{Output:} Reconstruction image $\bmcheckx$

	\begin{algorithmic}[1]
		\Function{ResiGLC Decoder}{$K, \mathbf{G}, \boldsymbol{r}$}

		\State $\bmhaty_M \gets \emph{ones}(N, C) \cdot \bm{m}_{\text{mask}}$ \Comment{Collect latent codes}
		\State $\mathbf{M} \gets \emph{ones}(N, C)$ 
		\If{$\boldsymbol{r}_{K+1} = 1$}
		\State $\boldsymbol{c} \gets \text{text\_decoder}(\mathcal{P}_{K+1})$  \Comment{Collect the text}
		\EndIf
		\For {$i \in \{1, \cdots, K\}$}
		\Comment{De-packetization}
		\If{$\boldsymbol{r}_i = 1$}
		\LineComment{Try to decode latent code slice $\mathcal{Y}_i$}
		\State \textbf{try:}
		\State \quad {$\y_{i, \text{ctx}} \gets$ \Call{ContextualModel}{$\mathcal Y, i, \mathbf{G}, \boldsymbol{r}$}}
		\If{any contextual code lost}
		\State $\boldsymbol{r}_i \gets 0$
		\State \textbf{continue}
		\EndIf

		\LineComment{All contexts received, decode slice $i$}
		\State $\bm{\mu}_i, \bm{\sigma}_i \gets h_{\text{entropy}}\left(f_{\text{MMT}}(\y_{i, \text{ctx}})\right)[\bm{n}_i, :]$
		\State $\mathbf{M}[\bm{n}_i, :] \leftarrow 0$ \Comment{Update mask}
		\State $\bmhaty_M[\bm{n}_i, :] \gets \text{entropy\_decoder}(\mathcal{P}_i, \bm{\mu}_i, \bm{\sigma}_i)$
		\EndIf
		\EndFor
		\LineComment{Stage I: Latent code loss concealment}
		\State $\check{\bm{y}} \gets h_{\text{PLC}}\left(f_{\text{MMT}}(\bmhaty_M)\right) \cdot \mathbf{M} + \bmhaty_M \cdot (\bm{1} - \mathbf{M})$
		\LineComment{Stage II: Conditional latent generation}
		\State Calculate the score estimate via~\eqref{eq_cfg}, and $\bm l_{t-1}$ via~\eqref{eq_ddim} in $T$ diffusion steps.
		\State $\bm{\check{l}}_0 \gets \boldsymbol{l}_{0|t=1}$
		\LineComment{Reconstruct image from the concealed token sequence}
		\State $\bmcheckx \gets \mathcal{D}(\bm{\check{l}}_0)$
		\State \Return reconstruction image $\check {\x}$
		\EndFunction
	\end{algorithmic}
	
\end{algorithm}

Specifically, we employ a latent diffusion model~\cite{rombach2022high} to characterize the latent generation. The generation flow $\bm{v}_t$ is a time-dependent function that transits from the Gaussian distribution to the target latent distribution.   
In the forward process, the noisy latent 
\begin{equation}\label{eq_fwd}
\boldsymbol{l}_t = \alpha_t\boldsymbol{l}_0 + \beta_t\bm{\epsilon},
\end{equation} 
where $\boldsymbol{l}_0 = \boldsymbol{l} = \mathcal{E}(\x)$, $\bm{\epsilon} \sim \mathcal N(\bm 0, \mathbf{I})$, and $(\alpha_t, \beta_t)$ is a predefined noise scheduler with $\alpha_t^2 + \beta_t^2 = 1, \alpha_0=1, \beta_0=0$.
The conditional generation starts from a standard Gaussian vector $\boldsymbol{l}_T$. 
The latent generator parameterized by $\theta$ predicts the velocity as $\bm{v}_t^{\theta}(\boldsymbol{l}_t, \boldsymbol{c}, \bmchecky)$, which is conditioned on the image caption and concealed latent codes. The goal is to regress the target velocity 
\begin{equation}
	\frac{d \boldsymbol{l}_t}{dt} = C(\alpha_t \bm{\epsilon} - \beta_t \boldsymbol{l}_0) \triangleq C\bm{u}_t,
\end{equation}
where $C$ is a constant.

During inference, we apply classifier-free guidance (CFG)~\cite{ho2021classifier} on the text condition. The resulting diffusion score is modified as the combination of the image-text conditioned score estimates and the image-only conditioned ones, as  
\begin{equation}\label{eq_cfg}
\bar{\bm{v}}_t^{\theta}(\boldsymbol{l}_t, \boldsymbol{c}, \bmchecky) = (w+1)\bm{v}_t^{\theta}(\boldsymbol{l}_t, \boldsymbol{c}, \bmchecky) - w\bm{v}_t^{\theta}(\boldsymbol{l}_t, \emptyset, \bmchecky),
\end{equation}
where $w$ is a positive parameter that controls the strength of the global textual semantic guidance.
In case of missing of the image caption, the receiver turns to condition solely on the concealed image latent codes, omitting classifier-free guidance.
In the backward sampling process, the predicted latent in each step is formulated by
\begin{equation}\label{eq_l0_pred}
\bm{l}_{0|t} = \alpha_t \bm l_t - \beta_t \bar{\boldsymbol{v}}_t^{\theta}.
\end{equation}
The latents for the previous timestep $t-1$ are derived via the DDIM update rule~\cite{song2021denoising}:
\begin{equation}\label{eq_ddim}
\begin{aligned}
\bm{l}_{t-1} &= \alpha_{t-1} \bm{l}_{0|t} + \beta_{t-1} \bm{\hat{\epsilon}}_t\\
&= \alpha_{t-1} \bm{l}_{0|t} + \beta_{t-1} \left(\beta_t \bm{l}_t + \alpha_{t}\bar{\bm{v}}_t^{\theta}\right).
\end{aligned}
\end{equation}
The resulting latents for pixel reconstruction come from the prediction in the final step $\bm{\check{l}}_0 \triangleq \bm{l}_{0|t=1}$.
The detailed workflow of ResiGLC receiver is illustrated in Algorithm~\ref{algorithm_receiver}.

In practice, image caption $\boldsymbol{c}$ is encoded by text semantic encoder and then incorporated into the generation via the cross attention mechanism. 
Latent codes are firstly rescaled to the same spatial dimension and then fused with the noisy token sequence $\boldsymbol{l}_t$ by channel-wise concatenation. Finally, they are mapped to condition vectors via a convolutional layer.

\begin{figure}[t]
	\setlength{\abovecaptionskip}{0.cm}
	\setlength{\belowcaptionskip}{-0.cm}
	\centering
	\includegraphics[width=\columnwidth]{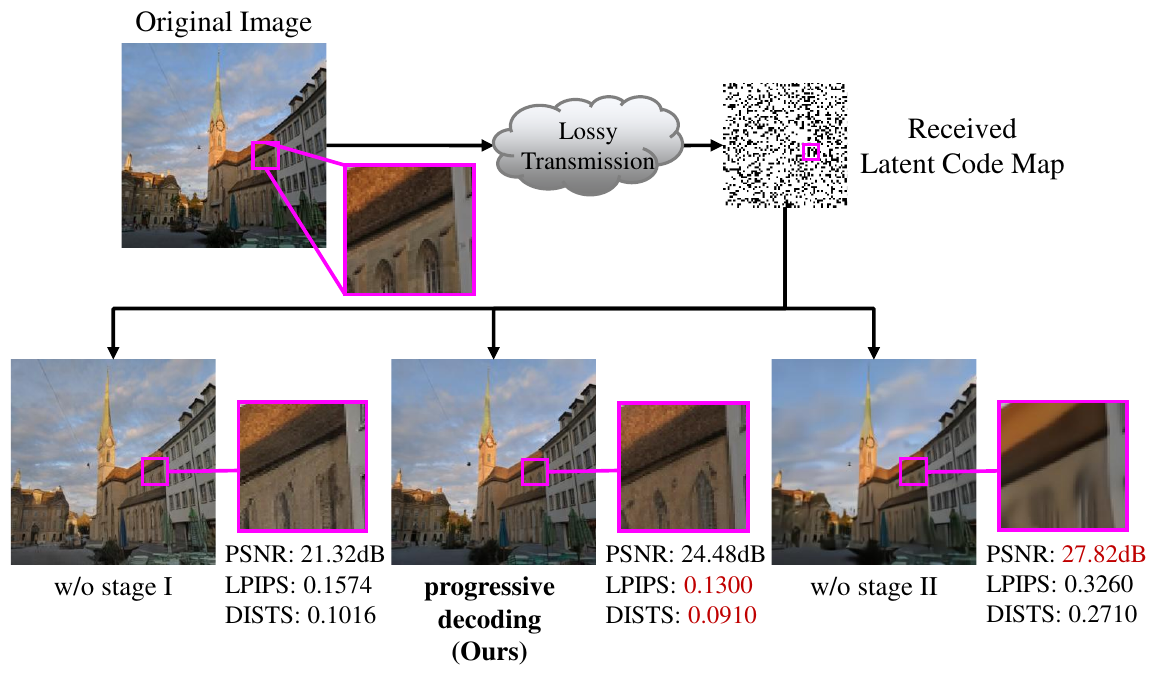}
	\caption{Visual comparison of different decoding schemes. \textbf{Left}: Generate the image latent $\bm{\check{l}}$ directly from $\bmhaty_M$ via a masked diffusion decoder. 
\textbf{Center}: ResiGLC's two-stage progressive decoding scheme.
\textbf{Right}: An image decoder reconstructs image from the restored latent codes $\bmchecky_M$. The average loss ratio of latent codes is $25\%$, and a binary latent code map is presented on the right top with black color indicating missing. The enlarged areas are marked by pink boxes, with spatial alignment in pixel space and latent code space. Please zoom in for better viewing.
	}\label{fig_resilient_comparison}
    \vspace{0em}
\end{figure}

Figure~\ref{fig_resilient_comparison} presents a visualization example directly comparing the reconstructions of different resilient decoding approaches.
We train a \textit{Masked Diffusion Decoder}, similar to~\cite{sahoo2024simple}, which omits the latent code concealment. 
We also provide another baseline that decodes image directly from concealed latent codes $\bmchecky$ (not in the realm of generative decoding). 
Particularly, another pixel decoder with Swin Transformer architecture~\cite{liu2021Swin} is employed, replacing the conditional diffusion model as well as the de-tokenizer. 
The reconstructions from the masked diffusion decoder (left) or decoding the latent codes (right) both suffer from severe visual artifacts and blurriness, especially in the regions with clustered loss (black area in latent code map).

\subsection{Training Strategy}

The training strategy of ResiGLC follows a three-phase training scheme, which progressively optimizes the conditional latent generator, the compression module, and the dual-functional MMT.

\subsubsection{Training text-conditioned latent generator}

In the first phase, we train the latent diffusion model assuming reliable transmission, i.e., the uncompressed latent codes $\y$ are available at the receiver. 
The tokenizer $\mathcal E$ is adapted from~\cite{rombach2022high} with a continuous generative latent space.
The neural backbone of the conditional latent generator is realized as a time-conditional UNet~\cite{unet}. $\boldsymbol{l}_t$ is efficiently obtained from $\mathcal E$ via~\eqref{eq_fwd} in the forward process, while the reconstruction image is decoded from the sample from $p(\boldsymbol{l})$ via the de-tokenizer $\mathcal D$. 
The training objective is to minimize the following loss:
\begin{equation}
  \mathcal{L}_G\left(t, \boldsymbol{c}, \y\right) =  \Vert \bm{u}_t(\boldsymbol{l}_t|\boldsymbol{l}_0) - \bm{v}_t^{\theta}(\boldsymbol{l}_t, \boldsymbol{c}, \y=\emptyset)\Vert^2,
\end{equation}
averaged over $t$ and data distribution $p_{\x}$.
$10\%$ random text dropping is implemented to enhance the robustness against the loss of textual condition.
In this phase, the diffusion model is not conditioned on the latent codes $\y$, which is set to the null condition $\emptyset$.

\begin{figure*}[t]
	\setlength{\abovecaptionskip}{0.cm}
	\setlength{\belowcaptionskip}{-0.cm}
	\centering
	\includegraphics[width=2\columnwidth]{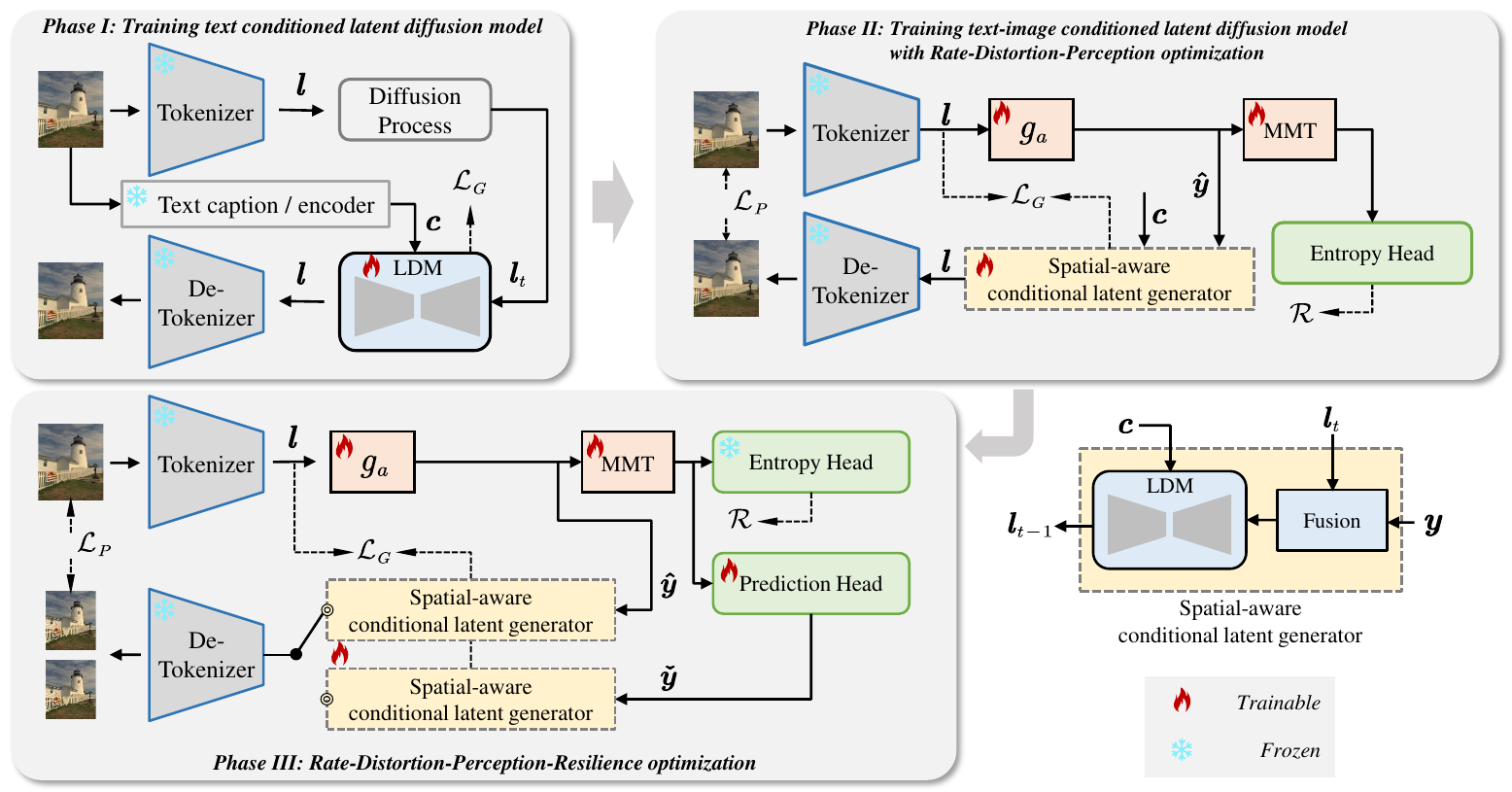}
	\caption{The progressive training procedure of ResiGLC consists of three phases. \emph{Phase I}: Train the text-conditioned latent diffusion model. \emph{Phase II}: Fine-tune the spatial-aware conditional latent generator conditioned on both the text prompt and the latent codes, and train the feature compressor $g_a$ and the MMT. \emph{Phase III}: Incorporate the MMT's prediction head, and jointly optimize the feature compressor, the MMT and the latent generator for rate-distortion-perception-resilience trade-off. 
	}\label{fig_train}
	\vspace{0em}
\end{figure*}

\subsubsection{Rate-Distortion-Perception optimization}

Assuming reliable transmission, the second training phase only optimizes the token sequence compression efficiency and effectiveness. 
The receiver has full access to the quantized latent codes $\bmchecky = \bmhaty = Q(\y)$ without the need of concealment.
We jointly optimize the compression module and the MMT for both compression efficiency performance. 
Particularly, a random binary mask $\mathbf{M}$ is applied on $\bmhaty$ to simulate \textit{arbitrary} context $\mathbf G$. The rate loss $\mathcal{R}$ is calculated on the masked latent codes as 
\begin{equation}
\mathcal{R} = - \sum_{i, \mathbf{M}_i = 1}\log_2 p(\bmhaty_i | \bmhaty_M).
\end{equation}

For better perceptual fidelity, we also use LPIPS~\cite{lpips} loss from the pixel reconstruction, which measures the reconstruction fidelity at the feature level by 
\begin{equation}
\mathcal{L}_{P}(\x, \bmcheckx)= \sum_l  {w_l} \Vert \psi _l(\boldsymbol{x})-\psi _l(\bmcheckx)\Vert _{2}^{2}.
\end{equation}
Here, $\bmcheckx$ is reconstructed from the predicted latent $\bm{l}_{0|t}$ in~\eqref{eq_l0_pred} for each sampled timestep $t$.
$\psi_l$ extracts features up to the $l$-th layer of a pre-trained VGG feature extractor and $w_l$ is the weight of the corresponding loss.

The goal of the second training phase is to achieve better compression performance, where the loss function is formulated as the rate-distortion-perception trade-off by
\begin{equation}
\mathcal L^{(2)} = \mathbb{E}_{\x \sim p_{\x},t\in(1,\cdots,T), \mathbf M} \left[\lambda \mathcal L_{G}\left(t, \boldsymbol{c}, \bmhaty\right) + \lambda_P \mathcal{L}_{P} + \mathcal{R} \right],
\end{equation}
where $\lambda, \lambda_P$ are hyperparameters that control the rate-distortion-perception trade-off.

\subsubsection{Rate-Distortion-Perception-Resilience optimization}

In the third training phase, we randomly mask out a part of latent codes to simulate the lossy transmission. 
The MMT is further fine-tuned to conceal the lossy latent codes through the prediction head, while maintaining its compression performance using the original quantized latent codes. 
The diffusion loss is calculated on both the quantized latent codes $\bmhaty$ and the concealed latent codes $\bmchecky$.
The overall loss function is summarized as
\begin{equation}
	\begin{aligned}
  \mathcal L^{(3)} &=  \mathbb{E}_{\x \sim p_{\x},t \in (1,\cdots,T), \mathbf M} \\ & \left[ \lambda \left(\mathcal L_{G}(t, \boldsymbol{c}, \bmhaty) +  \frac{\alpha}{1 + \alpha} \mathcal L_{G}(t, \boldsymbol{c}, \bmchecky) \right) + \lambda_P \mathcal{L}_{P}
		     + \mathcal{R} \right],
	\end{aligned}
\end{equation}
where the hyperparameter $\alpha > 0$ controls the efficiency and resilience trade-off.

The training pipeline is summarized in Fig.~\ref{fig_train}.
In practice, we vary $\lambda$ to obtain ResiGLC models at different coding rates, and set $\alpha=0.5$ to balance the resilience and efficiency.

\subsection{Packetization and Contextual Modeling Strategy}

The packetization scheme of ResiGLC plays a critical role in balancing coding efficiency and resilience against packet loss.
In addition to the text description which is transmitted in a dedicated packet, the encoded bitstream of $N$ latent codes with $C$ dimensions is divided into $K$ packets $\{\mathcal{P}_1, \cdots, \mathcal{P}_K\}$, resulting in a total of $K+1$ packets to be transmitted.

Next, we discuss the strategy for partitioning the latent codes into packets.
To enhance compression performance, we aim to minimize the conditional entropy $H(\mathcal{Y}_k|\mathcal{Y}_{\text{ctx}})$.
To achieve this, we establish our slice partitioner based on the quantized low-discrepancy sequences (QLDS) schedule~\cite{mentzer2023m2t}, which generates pseudo-random integer sequences that minimize the discrepancy among all subsequences.
Upon obtaining $K$ groups of latent codes, we apply the context map to define their dependencies. 
We evaluate three different context maps, including \textit{autoregressive} context, \textit{dual-autoregressive} context, and \textit{independent} context.
\begin{itemize}
\item
The autoregressive context defines a causal relationship among packets, i.e., the $k$-th packet can use all previous $k-1$ packets as contexts~(as plotted in Fig.~\ref{fig_MMT}, its context map $\mathbf{G}$ is a lower triangular matrix). 
\item The dual-autoregressive context divides the slices into two groups and defines autoregressive context relationship for them separately. 
\item The independent context indicates no contextual dependency, where $\mathbf{G}=\mathbf{I}_K$ is an identity matrix. 
\end{itemize}

The exploitation of intra-slice dependencies gradually decreases from the autoregressive to the independent context, leading to the improvement of resilience against packet loss but at the cost of reduced coding efficiency.
We observe that the choice of context maps has a distinct impact at a higher rate. Conversely, when emphasizing more on the rate term in model optimization, the impact on the coding efficiency gradually becomes marginal. 

\begin{figure}
	\setlength{\abovecaptionskip}{0.cm}
	\setlength{\belowcaptionskip}{-0.cm}
	\centering
	\includegraphics[width=1\columnwidth]{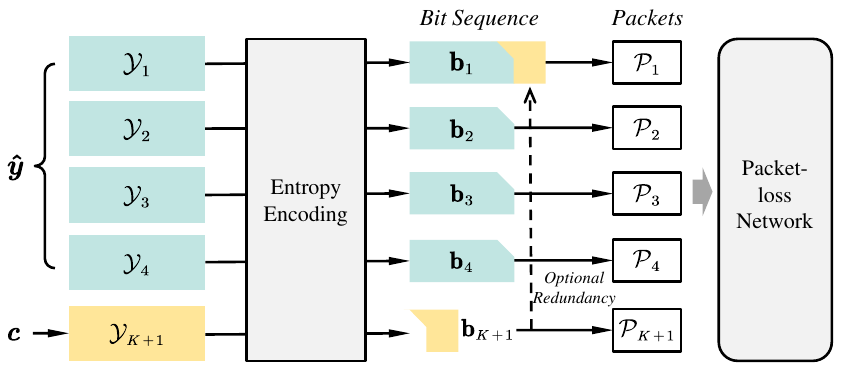}
	\caption{Slice-based packetization scheme tailored for autoregressive context. Text bitstream $\mathbf{b}_{K+1}$ is optionally appended in other non-contextualized packets ($\mathcal P_1$ here) as redundancy for prioritized protection of global semantics.
	}\label{fig_packetization}
	\vspace{-0.5em}
\end{figure}

Referring to the 5G NR standard~\cite{3GPP:TS38212}, and assuming a coding rate of $0.05 \sim 0.15$ bpp, the image data are segmented into multiple code blocks. 
We map each packet to one code block for error correction coding at the physical layer.
For simplicity, the packet number of ResiGLC is set to $K=4$ at $0.05-0.1$ bpp and $K=6$ for $>0.1$ bpp. 
As shown in Fig.~\ref{fig_packetization}, FEC redundancy for text can be appended in other packets, to ensure prioritized protection of text semantics. Particularly, the packets that do not contextually depend on any packet are allowed to carry redundancy. 
{In extreme cases where none of the packets are received, the decoder has no choice but to request retransmission since having no access to any available information of the image.}
The bandwidth overhead (estimated to be as less as $0.002$ bpp equivalently) for the text redundancy, is negligible compared to that for latent codes.
For simplicity, we assume that packet loss can be accurately detected at the receiver, via the packet headers or code-block-level cyclic redundancy checking. In other words, the reception state of latent codes is known to the receiver.

\section{Experiments}\label{section_experiments}

\subsection{Experimental Setup}

\textit{ 1) Datasets and implementation details}:
ResiGLC is trained on OpenImages~\cite{openimages} training set.  
All images are resized to $256\times256$ resolution, and randomly flipped horizontally for training. 
ResiGLC model is trained with the Adam optimizer with a batch size of $8$ and learning rate of $10^{-4}$.
Evaluations are conducted on $512\times512$ images of CLIC2020 dataset~\cite{clic2020} and $512\times768$ images of Kodak dataset~\cite{kodak}.
The channel dimension of latent codes $C$ is $320$ for coding rate greater than $0.1$ bpp and $160$ otherwise.

\begin{figure}
	\setlength{\abovecaptionskip}{0.cm}
	\setlength{\belowcaptionskip}{-0.cm}
	\centering
	\includegraphics[width=0.75\columnwidth]{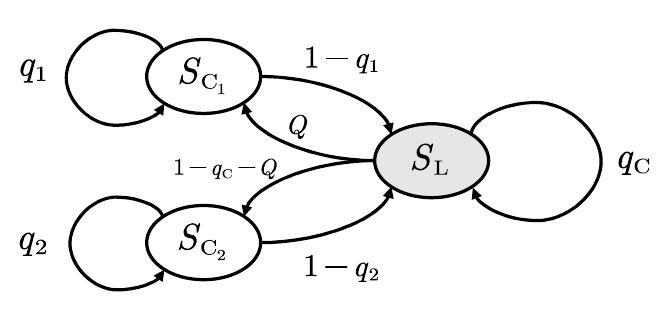}
	\caption{State transfer diagram of three-state Markov channel model.
	}\label{fig_markov_channel}
	\vspace{0em}
\end{figure}

\textit{ 2) Packet-loss channel setup}: We conduct our simulations on memoryless packet lossy channels with a given loss ratio and channels with memory. The packet loss trace of memory channel is simulated by a three-state Markov model~\cite{markovchannel}.
The state transfer characteristics are described in Fig.~\ref{fig_markov_channel}, comprising the two loss-free states $S_{\text{C}_1}$ and $S_{\text{C}_2}$ with packet correctly received, and one state $S_{\text{L}}$ denoting packet loss.
The Markov model is widely used and demonstrated effective in reproducing the characteristics of real network traces, especially for burst-like conditions with consecutive packet loss.
The parameter $q_\text{C}$ depicts the probability of consecutive packet loss. Allowing independent self-loop probabilities $q_1$ and $q_2$ for two loss-free states, the model enables two types of error-free periods.
Packet-level error patterns (EP) are generated to simulate wireless local area network (WLAN) channel (EP1) and wireless cellular networks (EP2 and EP3 simulate GSM channels)~\cite{milner2004packet}, and the corresponding parameters are shown in Table~\ref{tab:markov}.

\begin{table}
\centering
\caption{Parameter set $\kappa$ of 3-state Markov channel models}
\label{tab:markov}
\normalsize
\begin{tabular}{c|c|c|c|c}
\hline
Channel model &  $q_\text{C}$ & $q_1$ & $q_2$ & $Q$  \\
\hline
EP1 (WLAN) & 0.4072 & 0.9363 & 0.5652 & 0.3631 \\
EP2 (GSM) & 0.2742 & 0.9400 & 0.5821 & 0.6977 \\
EP3 (GSM) & 0.6305 & 0.8508 & 0.2000 & 0.2982 \\
\hline
\end{tabular}
\end{table}

\textit{ 3) Comparison schemes}: We compare our ResiGLC with several representative traditional and neural image codecs, including BPG~\cite{bpg}, HiFiC~\cite{mentzer2020high}, MS-ILLM~\cite{msillm}, PerCo~\cite{careil2023towards} and DiffEIC~\cite{diffeic}. 
To evaluate the resilience, these codecs, combined with FEC for protection, are denoted as ``DiffEIC + $x$\% FEC'' for example.
We also compare with the state-of-the-art resilient image codecs LRIC~\cite{sha2025towards} and ResiComp~\cite{wang2025resicomp}.

\subsection{Metrics}

We employ established metrics to assess the rate-distortion-perception performance of ResiGLC. 
Concretely, the total bandwidth is measured in equivalent bpp, which accounts for the coding bits of latent codes and text description. For non-resilient codecs, FEC bitrate cost is also included. 
We exclude other bandwidth cost, including error-correction codes in physical layer and other protocol overheads, as they remain consistent for all evaluated methods.

Objective fidelity is assessed by peak signal-to-noise ratio (PSNR), which equals $-10 \log_{10} \text{MSE}(\x, \bmhatx)$.
We evaluate perceptual quality using FID~\cite{heusel2017gans}, KID~\cite{binkowski2018demystifying}, DISTS~\cite{ding2020image} and LPIPS~\cite{lpips}.
Among them, FID and KID are used to evaluate perceptual realism by matching the feature distributions between the original and reconstructed image sets. They are calculated on $256\times256$ overlapping patches, as described in HiFiC~\cite{mentzer2020high}.
The latter two metrics are reference-based metrics measuring the perceptual fidelity by calculating the distance of deep features of two images. 
In the following figures, the upward arrow ``$\uparrow$'' indicates that higher values of the metric are favorable, and vice versa.

\subsection{Resilience Performance}

\begin{figure*}
	\setlength{\abovecaptionskip}{0.cm}
	\setlength{\belowcaptionskip}{-0.cm}
	\centering
	\includegraphics[width=2\columnwidth]{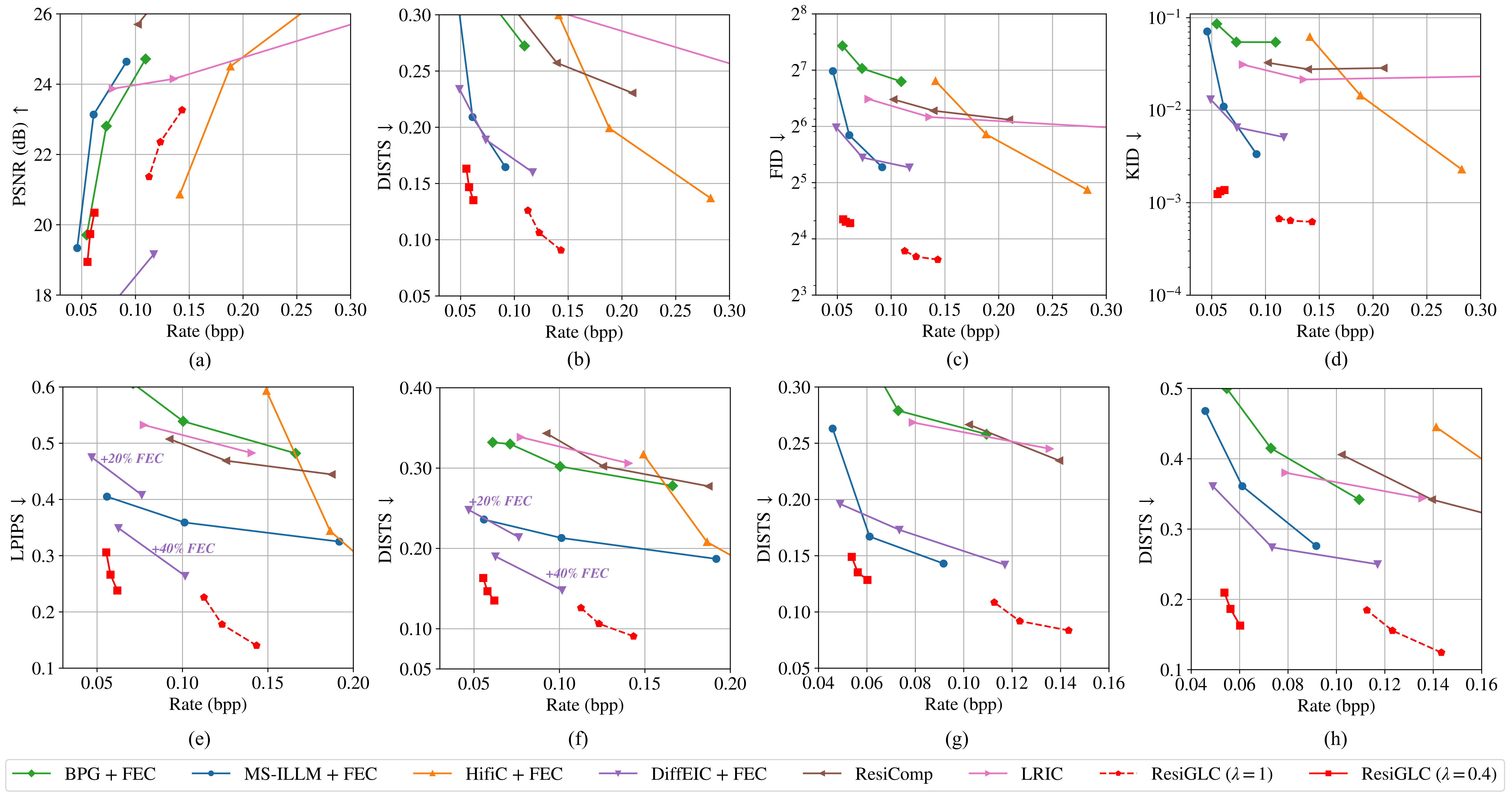}
	\caption{Resilience performance over three-state Markov channels. 
	(a--d) Fidelity and realism comparison on resized $512\times512$ images from CLIC2020 dataset over the WLAN channel (EP1).
(e--f) LPIPS and DISTS metrics evaluated on Kodak dataset at original resolution under EP1.
(g--h) DISTS performance on the CLIC2020 dataset over GSM channels EP2 and EP3, respectively. 
	}\label{fig_resi_markov}
	\vspace{-0.3em}
\end{figure*}

We conduct packet-level simulations to evaluate the rate-distortion-perception performances over three-state Markov channels in Fig.~\ref{fig_resi_markov}. 
Traditional codecs reliant on FEC redundancy are highly susceptible to bursty loss. 
We add appropriate FEC based on the average loss ratio of the packet loss trace. In the event of decoding failure (i.e., when burst loss exceeds the FEC capability), a random noise image is substituted for performance evaluation.
In contrast, ResiGLC avoids an abrupt quality degradation provided that any packet is decoded. 
The results, averaged over extensive simulations, demonstrate that ResiGLC consistently outperforms competing methods in perceptual quality, while maintaining comparable PSNR. 
For ResiGLC, points on each curve are obtained using different context maps $\mathbf G$ based on the same model. 
This illustrates that by selecting different context modes, ResiGLC effectively navigates trade-offs between error resilience and compression efficiency.  
The trend is particularly pronounced in the lower bitrate regime, characterized by a steeper slope. 
Specifically, Figs.~\ref{fig_resi_markov}(a)-(f) display the performance over WLAN channels, 
while Figs.~\ref{fig_resi_markov}(g) and (h) further present the resilience performance under two error patterns simulating GSM channels. 
The misalignment of FEC ratios with exact packet loss traces is emphasized under longer burst errors. 
We also provide the performance of perceptual image codec with high FEC redundancy (20\% and 40\% FEC), as shown in Figs.~\ref{fig_resi_markov}(e) and (f). Despite increased decoding success probability, it sacrifices the effective image coding rate, therefore compromising bandwidth efficiency.

By contrasting DISTS in Fig.~\ref{fig_resi_markov}(b) with FID/KID metrics in Figs.~\ref{fig_resi_markov}(c) and (d), we observe that enhancing packet-loss resilience of latent codes contributes more notably to perceptual fidelity than to realism. 
This suggests that ResiGLC secures a baseline of image naturalness from text-conditioned generative prior, and then image conditions serve primarily to refine structural alignment.
Despite the marginal gain in realism, it consistently outperforms existing perceptual generative image codecs, validating its capability for resilient generative latent decoding from incomplete conditions.

\begin{figure}
	\setlength{\abovecaptionskip}{0.cm}
	\setlength{\belowcaptionskip}{-0.cm}
	\centering
	\includegraphics[width=\columnwidth]{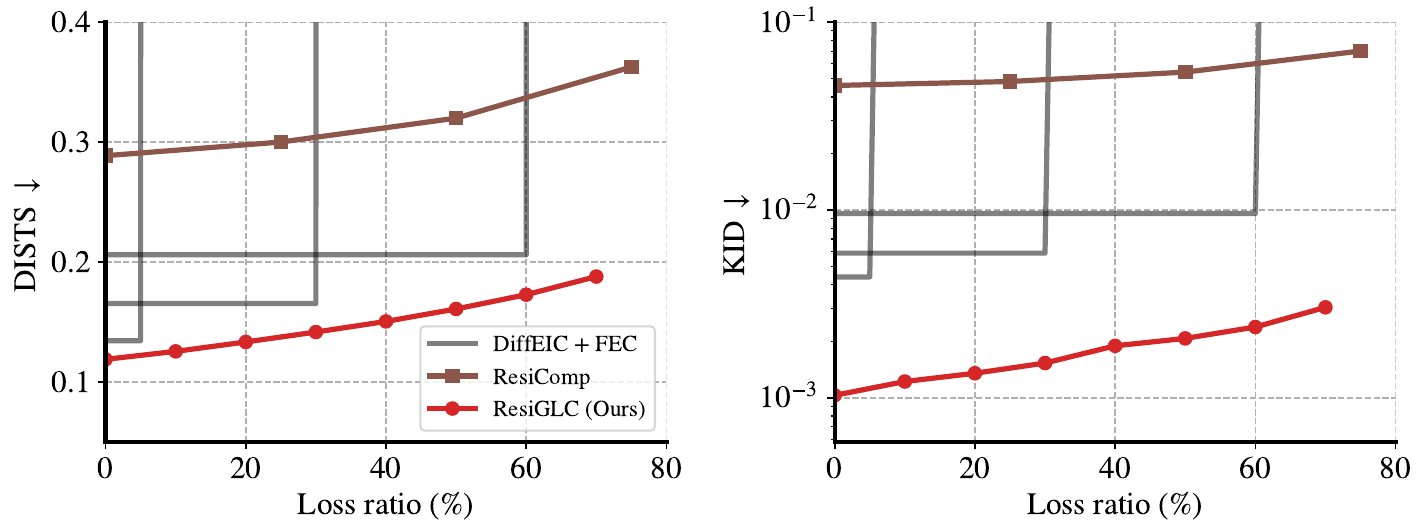}
	\caption{Loss resilience in terms of perceptual fidelity and realism under different loss ratios. The rate for all schemes is constrained below 0.06 bpp.
	}\label{fig_dmr}
    \vspace{0em}
\end{figure}

Results over memoryless packet lossy channels are presented in Fig.~\ref{fig_dmr}, where we evaluate perceptual fidelity (DISTS) and realism (KID) across various loss ratios, with the total rate constrained below 0.06 bpp. ResiComp exhibits inferior performance in both aspects, as it is optimized toward lower objective distortion only. While perceptual image codec DiffEIC offers sufficient perceptual quality in reliable transmission cases, its robustness relies on the FEC redundancy. In contrast, ResiGLC demonstrates inherent resilience to packet loss, attributed to the preliminary concealment of conditioning signals prior to token generation.

\subsection{Semantic Consistency}\label{subsec_res_sem}

\begin{figure}
	\setlength{\abovecaptionskip}{0.cm}
	\setlength{\belowcaptionskip}{-0.cm}
	\centering
	\includegraphics[width=0.94\columnwidth]{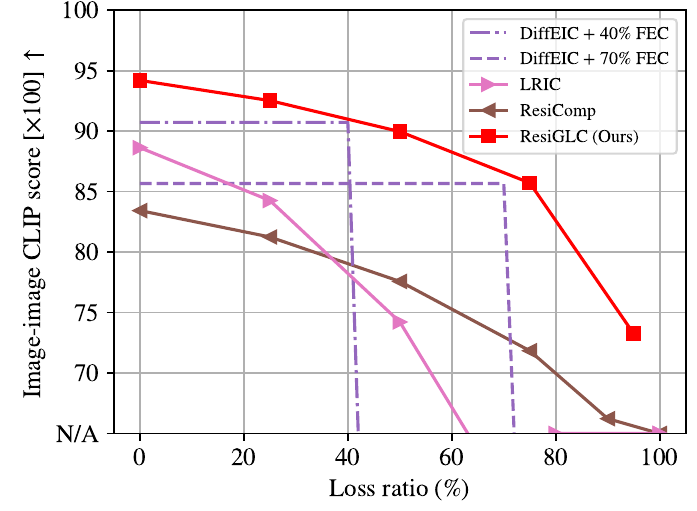}
	\caption{
		Semantic consistency evaluation in terms of CLIP score vs. the loss ratio of latent codes.  
	``N/A'' indicates either the image cannot be decoded correctly or the value is below the minimum.
	The total coding rate budget is 0.065 bpp for $512 \times 512$ images, including possible FEC redundancy.
	}\label{fig_clip}
	\vspace{0em}
\end{figure}

CLIP score measures global alignment via cosine similarity between the CLIP embeddings~\cite{hessel2021clipscore}. We first assess the CLIP score between the original image and the reconstructed one. The results in Fig.~\ref{fig_clip} highlight the outstanding consistency across all loss ratios. With the increase of loss ratio, the traditional codecs reach their operational limits given the bandwidth budget, as sufficient FEC redundancy is required for successful decoding.
While resilient image codecs like ResiComp and LRIC, present slower performance drop compared to ``\textit{codec} + FEC'' methods, the semantic consistency deteriorates significantly as latent code loss intensifies. 
Conversely, ResiGLC still maintains high semantic consistency in the high loss regime, which is attributed to the two-stage resilient coding utilizing both the latent code prior and the text-conditioned diffusion prior.

\begin{table}
\centering
\caption{Image-text CLIP score versus the loss ratio of latent codes}
\label{tab:t2i_clip}
\normalsize
\renewcommand{\arraystretch}{1.2}
\begin{tabular}{c|c|c|c|c|c}
\hline
Loss ratio (\%) & 0 & 25  & 50 & 75 & 95 \\
\hline
CLIP score $\uparrow$ & 29.4 & 29.6 & 29.9 & 30.4 & 31.2 \\
\hline
\end{tabular}
\end{table}

We also evaluate the image-text CLIP score to measure the global textual semantic preservation under packet lossy channels, i.e., the similarity between the global feature vector of the reconstructed images and that of the image caption. 
Table~\ref{tab:t2i_clip} shows the CLIP score comparison under different loss ratios of latent codes. 
Counterintuitively, despite the same text-conditioning strength and diffusion steps $T=10$, the alignment with textual semantics is better even with the increase of loss ratio. 
This phenomenon indicates the conditional latent generator intrinsically shifts its dependency: as more latent codes are lost and concealed by MMT, the generator learns to rely more on the text conditions.  
Please refer to Subsection~\ref{subsec_visual} for visualization.

\subsection{Impact of Text Conditioning Strength}

\begin{figure}[htbp]
	\setlength{\abovecaptionskip}{0.cm}
	\setlength{\belowcaptionskip}{-0.cm}
	\centering
	\includegraphics[width=0.95\columnwidth]{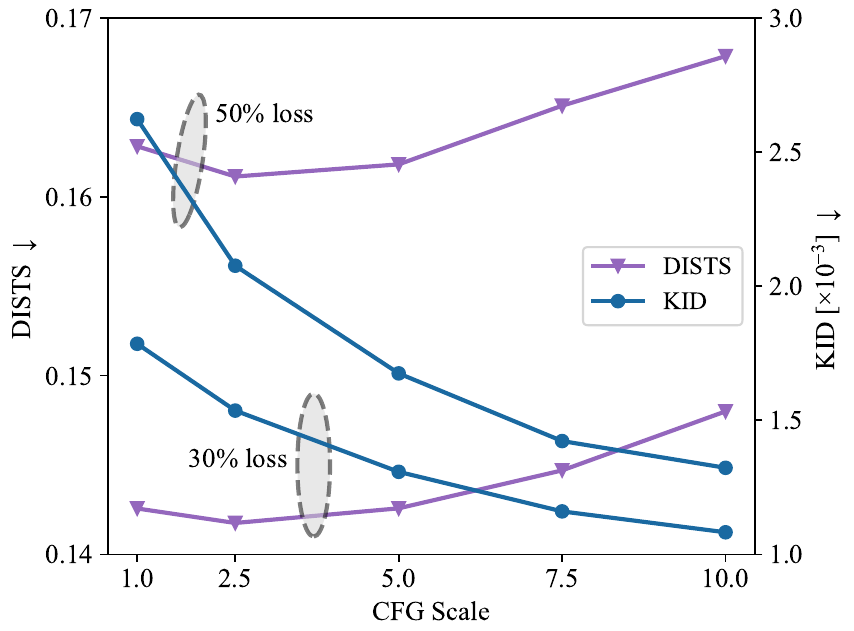}
	\caption{
		Fidelity and realism performances across different text conditioning scales when dropping 30\% and 50\% image latent codes at the receiver end.
	}\label{fig_cfg}
	\vspace{0em}
\end{figure}

As the global text conditioning is adjusted subject to availability, we additionally investigate the impact of the conditioning strength of the text condition within the CFG inference framework.
Results in Fig.~\ref{fig_cfg} show that KID decreases with stronger text-conditioning strength. This indicates better naturalness when leveraging more global semantics as well as the diffusion prior. 
Conversely, the fidelity is better preserved with reduced text guidance. 
This phenomenon is more pronounced at higher loss ratios, where a larger scale is preferred to improve the realism. The primary reason is that the MMT concealment in stage I becomes more challenging and the resulting spatial image conditioning in stage II is not accurate, which fails to contribute to and may even degrade the realism of the reconstruction. 
In practice, we adopt a CFG scale of $w \in [2.5, 7.5]$.

\begin{table}[htbp]
\centering
\normalsize
\caption{Comparison of different image captioning methods under $10\%$ packet loss}
\label{tab:ablation_study_caption}
\resizebox{\columnwidth}{!}{
\begin{tabular}{c|c|c|c|c}
\toprule
\textbf{Text Type} &  \makecell{\textbf{Image-text} \\ \textbf{CLIP Score} $\uparrow$} & \textbf{FID} $\downarrow$ & \textbf{KID} $\downarrow$ &  \makecell{\textbf{Time Cost} \\ \textbf{(ms)}} \\
\midrule
BLIP-2      &  29.5  & 19.3 & 0.00134 & 295 \\
OFA-Tiny     &  26.4  & 20.9 & 0.00179 & 96 \\
Empty string   & 28.5 & 20.1 & 0.0022 & 0 \\
\bottomrule
\end{tabular}
}
\end{table}

We evaluate three different approaches to evaluate the robustness regarding the text caption, including (1) BLIP-2~\cite{li2023blip}, a visual-language model which consists of a CLIP-like encoder and a querying Transformer; (2) OFA-Tiny, a lightweight multi-modal model OFA~\cite{wang2022ofa} with 33M parameters; (3) Empty string, i.e., without encoding and transmitting image caption. The decoder is conditioned solely on the image tokens.

Table~\ref{tab:ablation_study_caption} compares the averaged reconstruction quality under $10\%$ packet loss, showing that both semantic consistency and perceptual realism are improved with the quality of image description.
The text conditioning is randomly dropped during training, which enhances the robustness of ResiGLC against the loss of text description. The ResiGLC decoder can generate images solely conditioned on the received image tokens. 
The time cost of image captioning varies. Using OFA-Tiny model reduces the time cost from $295$ms to $96$ms for $512\times512$ images, at the expense of image quality. Note that ResiGLC's decoder takes a short description ($\sim 10$ words) as input, the time cost for image captioning is not significant.

\begin{figure*}[htbp]
	\setlength{\abovecaptionskip}{0.cm}
	\setlength{\belowcaptionskip}{-0.cm}
	\centering
	\includegraphics[width=2\columnwidth]{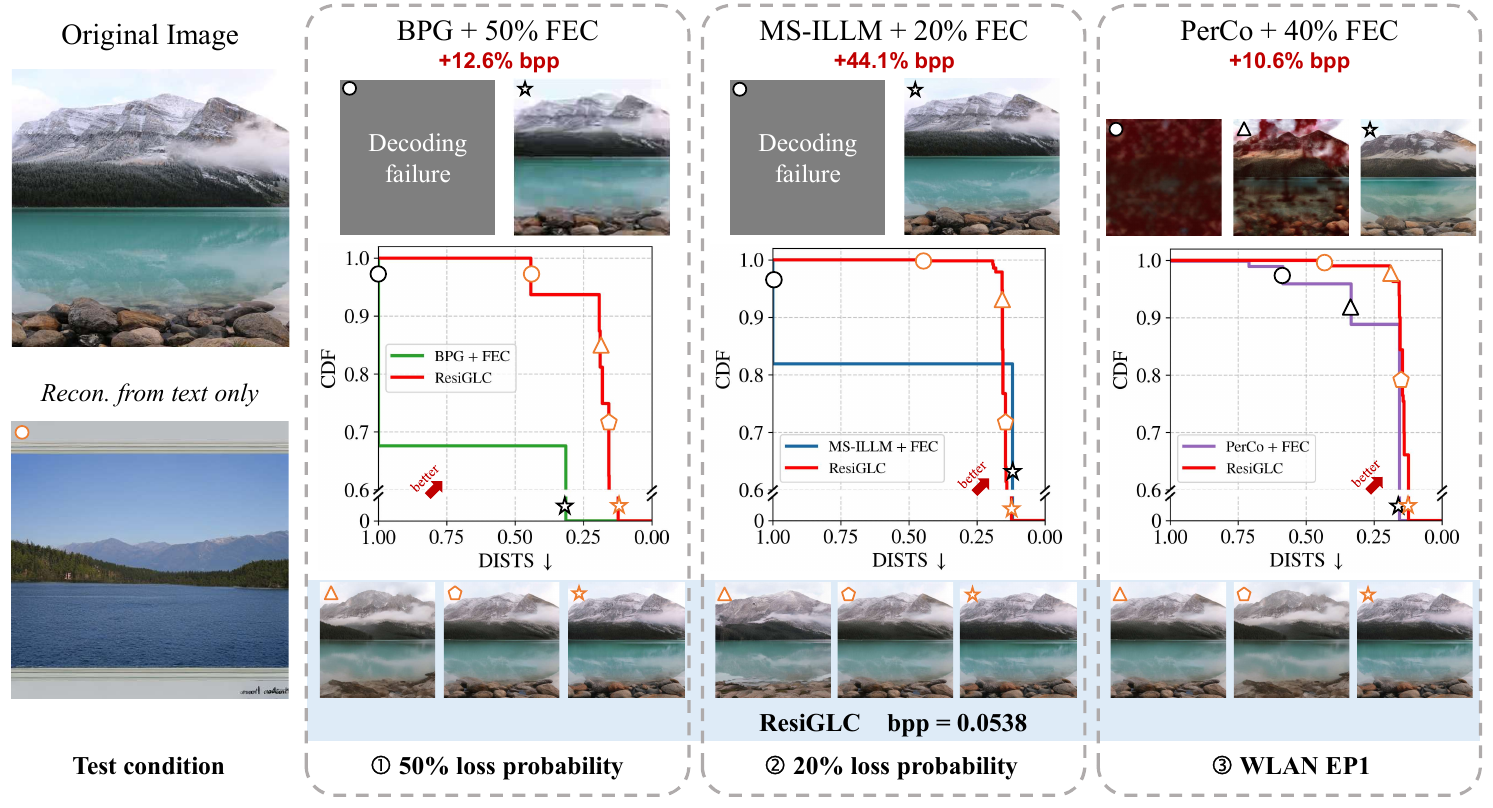}
	\caption{Visualization example of resilient image transmission over packet loss channels. 
	The second and third columns correspond to memoryless packet loss probability of 50\% and 20\%, and the fourth column corresponds to WLAN channel (EP1). 
	The cumulative distribution functions (CDFs) of DISTS performance are reported, where the curve closer to the upper right is better. 
	Crucially, the distinct markers (e.g., $\star, \circ, \triangle$) on the CDF curves correspond to the specific reconstructed images displayed in the top and bottom rows.
	For existing image codecs (top row), the FEC redundancy ratio is chosen based on channel characteristics. The coding rate including FEC is marked in red, anchored by ResiGLC. 
	As the practical packet loss fluctuates, the image reconstruction may fail (indicated by ``$\circ$'' on the CDF curves), attributing to the entropy decoding failure. 
	Instead, ResiGLC (bottom row) can reconstruct images with varying numbers of packets, exhibiting robust communication even without knowledge of channel statistics. 
	The image is generated solely conditioned on the caption ``\textit{A large body of water with a mountain in the background}'' when only text is received (bottom leftmost).
	}
	\label{fig_resi_visual}
	\vspace{0em}
\end{figure*}

\subsection{Reconstruction from Deconstruction}

The denoising paradigm of modern generative models has inspired numerous works in visual tokenizers~\cite{He_2022_CVPR}.
Typical noising strategies take place in latent space including masking noise (apply a binary mask) and Gaussian noise, which can be collectively characterized as structured noise with analytical probabilistic distribution function.  
They share the conceptually similar objective, i.e., to reconstruct original signals from deconstructed ones.
However, the ``noise'' from latent codes over lossy channels departs from the denoising mode of these generative models, which is the major obstacle that prevents them from generalizing to channel impairments.

In synergy with MMT, ResiGLC reformulates the denoising latent diffusion models to adapt to the disturbance of latent codes. 
The conditional latent generator in decoding stage II removes diffusion-induced noise in generative latent space to recover clean latents, but also learns the generation from noisy condition signals in the latent code space.

\begin{table}[htbp]
\centering
\caption{Ablation study of two-stage resilient decoding.}
\label{tab:ablation}
\normalsize
\begin{tabular}{c|c|c|c}
\hline
Model & PSNR (dB) $\uparrow$ & DISTS $\downarrow$ & FID $\downarrow$ \\
\hline
ResiGLC & 19.74 & \textbf{0.146} & \textbf{19.76} \\
w/o stage I & 17.89 & 0.177 & 33.65 \\
w/o stage II & \textbf{24.39} & 0.324 & 106.88 \\
\hline
\end{tabular}
\vspace{0em}
\end{table}

We examine the effectiveness of two-stage resilient decoding under the same Markov channel setting, as shown in Table~\ref{tab:ablation}.
Latent code concealment in stage I improves both fidelity and realism when employing generative latent features to reconstruct images. 
We also train a baseline model at similar coding rate that directly decodes images from $\bmchecky$ via another pixel decoder (``w/o stage II'' in the table). Despite better pixel-level fidelity, the perceptual quality and realism are undermined in this bitrate regime.
A visualization example can be found in Fig.~\ref{fig_resilient_comparison}.

\subsection{Visualization}\label{subsec_visual}

\begin{figure*}[ht]
	\setlength{\abovecaptionskip}{0.cm}
	\setlength{\belowcaptionskip}{-0.cm}
	\centering
	\includegraphics[width=2\columnwidth]{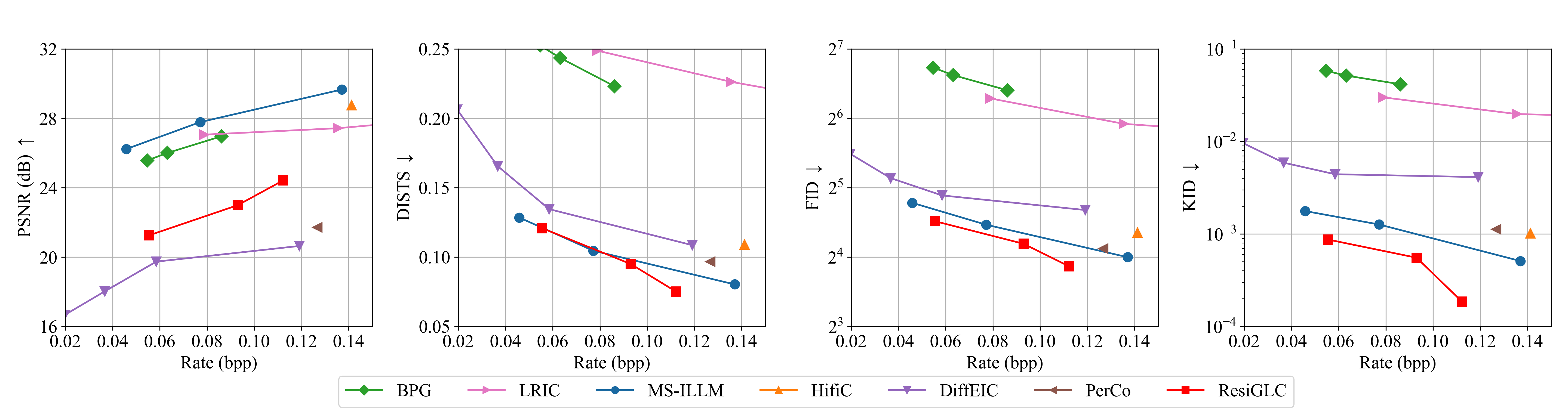}
	\caption{Rate-distortion-perception compression performance on CLIC2020 dataset compared with traditional and perceptual image codecs.
	}\label{fig_rd}
	\vspace{-0.5em}
\end{figure*}

\begin{figure*}[ht]
	\setlength{\abovecaptionskip}{0.cm}
	\setlength{\belowcaptionskip}{-0.cm}
	\centering
	\includegraphics[width=2\columnwidth]{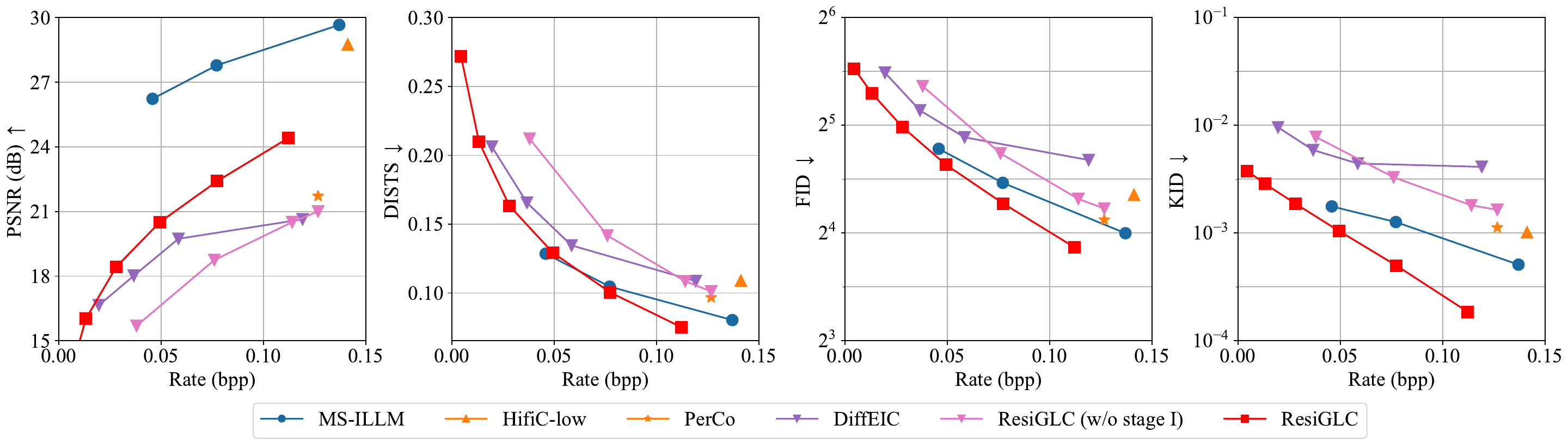}
	\caption{Rate-distortion-perception performances in progressive coding cases. ResiGLC is evaluated using the same model, where the rightmost point indicates the full coding rate, sending all the latent code slices.
	}\label{fig_prog}
	\vspace{-1em}
\end{figure*}

Figure~\ref{fig_resi_visual} provides an intuitive comparison of ResiGLC's resilience under different packet loss patterns, including memoryless packet loss probabilities (50\% and 20\%) as well as a bursty WLAN channel (EP1). 
In addition to averaged metrics, we report the cumulative distribution functions (CDFs) of DISTS performance to visualize the variance in reconstruction quality.

For traditional codecs, the FEC ratio is configured by the average loss ratio, and is redundant in WLAN channel. Despite this, the image reconstruction may fail as the packet loss fluctuates, attributed to the breakdown of entropy decoding. For PerCo, which employs vector quantization instead of entropy coding, the lost indices are substituted with zero values before decoding. 
Traditional codecs have to rely on stringent link adaptation and feedback to the sender for FEC configuration, which is hindered in extreme-bandwidth scenarios. 
On the contrary, ResiGLC obtains reconstructions with more stable objective and perceptual fidelity, even without prior knowledge of channel statistics at the sender side. 
This property underscores the flexibility that the visual communication system can simplify the link adaptation and other error control mechanisms, enabling intrinsically robust semantic communications in bandwidth-limited scenarios.

\subsection{Rate-Distortion-Perception Compression Performance}

The rate-distortion-perception compression performance on CLIC2020 dataset is shown in Fig.~\ref{fig_rd}. BPG, MS-ILLM, HIFI-C, and DiffEIC are tested using their officially released models. BPG is tested using Quantizer Parameter $\mathrm{QP}=48, 50$, and $51$, respectively. 
HIFI-C is tested using the \textit{hific-low} profile.
ResiGLC is evaluated on models with $\lambda=0.4, 1.0$, separately. 
We observe that ResiGLC achieves similar or better perceptual quality (fidelity and realism) below $0.1$ bpp, compared to traditional and neural image codecs. 
While a gap still appears in PSNR performance compared to non-generative image codecs (e.g., MS-ILLM and HIFI-C), ResiGLC apparently excels among the existing diffusion-based image codecs, like DiffEIC~\cite{diffeic} and PerCo~\cite{careil2023towards}.

ResiGLC divides the latent codes into several slices and can decode images using a subset of them.  
It intrinsically supports the feature of progressive coding, benefiting from the context modeling with arbitrary context map and the resilient decoding. 
Fig.~\ref{fig_prog} shows the performance of progressive coding and transmission over reliable channel. 
Specifically, we anchor our ResiGLC model at 0.112 bpp as the full coding rate model, where the latent codes are divided into $K=6$ slices. 
The red curve illustrates the quality improvement with more latent codes received, in terms of less objective distortion and better perceptual quality.
Over reliable channels, we emphasize the potential of ResiGLC to flexibly fit the channel bandwidth to determine which and how many latent code slices to send.

\subsection{Complexity Analysis}

\begin{table}[t]
		\centering
		\caption{
			Computational complexity and averaged encoding/decoding latency comparison
		}
		\tabcolsep=0.18cm
		\footnotesize
		\resizebox{\linewidth}{!}{
			\begin{threeparttable}
				
				\begin{tabular}{m{1.7cm}<{}m{1.0cm}<{\centering}m{0.3cm}<{}m{1.1cm}<{\centering}m{1.1cm}<{\centering}m{1.1cm}<{\centering}m{0.6cm}<{\centering}}
					\toprule
					\multirow{2}{*}{Method} & \multirow{2}{*}{\parbox[c]{0.6cm}{Diffusion\\model}} & \multirow{2}{*}{$K$} & \multicolumn{2}{c}{Inference Time $^*$} & \multirow{2}{*}{Param. $^{\ddag}$} \\
					\cmidrule(lr){4-5}
					&&& {Enc.} & {Dec.}  \\
					\midrule
					GMM-Attn~\cite{cheng2020learned} & \multirow{2}{*}{$\times$} & $N$ & $78$ms & $>10^3$ms & $20$M \\
					ResiComp~\cite{wang2025resicomp} && $10$ & $131$ms & $167$ms & $128$M \\
					\midrule
					DiffEIC~\cite{diffeic} & \multirow{3}{*}{$\checkmark$} & $20$ & $310$ms & $>10^3$ms  & $1379$M \\
					PerCo~\cite{careil2023towards} && N/A & $46$ms & $689$ms & $955$M \\
					\emph{ResiGLC} && $4$ & $36$ms & $615$ms  & $969$M \\   
					\bottomrule
				\end{tabular}
				
				\begin{tablenotes}
					\footnotesize
					\item[*] All experiments are conducted on the same platform with an Intel Xeon Gold 6226R CPU, an RTX 4090 GPU, with PyTorch 2.3.0 and CUDA 11.8. Inference time includes the model inference time and the entropy enc./dec. time. All diffusion models use $T=10$ steps. 
					\item[$\ddag$] The image captioning and text encoder modules are excluded for the parameter count. 
				\end{tablenotes}
			\end{threeparttable}
		}
		\label{tab:complexity}
		\vspace{-1em}
	\end{table}

We report the computational and space complexities compared to existing image codecs in Table~\ref{tab:complexity}.
We measured the encoding and decoding time on GPU, including the model inference time and the entropy encoding/decoding time, under the packet-loss scenarios. 
The parameter $K$, denoting the number of slices that the compressed vectors are partitioned into, is also listed. For instance, DiffEIC uses $20$ slices, comprising $10$ channel-wise and $2$ spatial splits. 
As observed, non-diffusion-based image codecs generally maintain an advantage in latency, without the multi-step diffusion sampling.
ResiGLC's two-stage resilient decoding does not incur significant extra computational overhead among diffusion-based models. 
In terms of space complexity (measured in parameter count), the MMT module in ResiGLC is lightweight compared to diffusion backbone itself.

\begin{figure}[htbp]
    \begin{minipage}[t]{0.28\textwidth}
		\vspace{0pt}
        \centering
        \includegraphics[width=\textwidth]{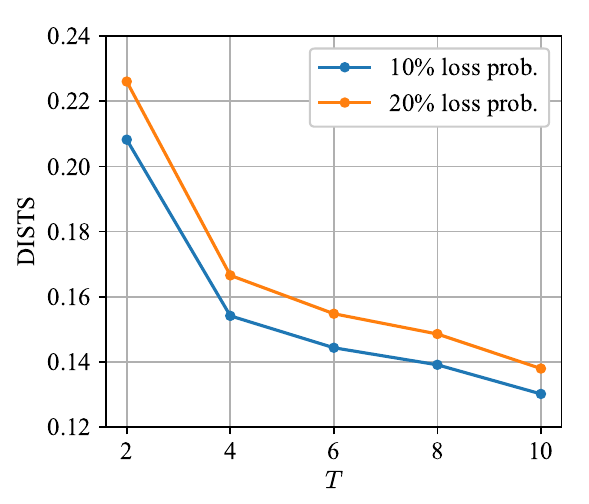}
    \end{minipage}%
	\hspace{-0.5em}
	\begin{minipage}[t]{0.15\textwidth}
    \vspace{0.8em}
    \small
    \centering
    \renewcommand{\arraystretch}{1.3}
    \begin{tabular}{c|cc}
        \toprule
        \multirow{2}{*}{$T$} & \multicolumn{2}{c}{Decoding time (ms)} \\
        \cmidrule(lr){2-3} & 4090 & 2080Ti \\
        \midrule
        2  & 150 & 442 \\
        4  & 265 & 853 \\
        6  & 379 & 1315 \\
        8  & 492 & 1721 \\
        10 & 615 & 2160 \\
        \bottomrule
    \end{tabular}
\end{minipage}

  \caption{Quality-latency trade-off curves under different numbers of diffusion steps $T$. $K=4$ slices at $\sim0.06$ bpp for $512\times512$ images are encoded and transmitted over i.i.d. packet loss channel with 10\% and 20\% loss probability each.
  Decoding latencies are measured on one RTX 4090 and RTX 2080Ti GPU, respectively.}
  \label{fig:quality_step}
\end{figure}

We compare the performance of ResiGLC with fewer diffusion steps $T$ in Fig.~\ref{fig:quality_step} to investigate the trade-off between the quality and decoding latency. 
The latent generation in stage II remains the critical bottleneck for practical deployment, especially for weaker GPUs. 
Other pipelines including entropy decoding and concealing in stage I take about 40 ms. 
Too few diffusion steps ($T=2$) prevent sufficient feature fusion with concealed latent codes. As $T$ increases to $10$, the perceptual quality improves significantly, while the decoding latency increases by more than $4$ times. In practice, a smaller $T$ can be adopted to achieve a better quality-latency trade-off, to meet diverse latency requirements and hardware constraints.

\section{Conclusion and Future Work}\label{section_conclusion}

This paper proposes \textit{ResiGLC}, a loss-resilient generative semantic communication framework. 
Through progressive decoding, it achieves substantial resilience in perceptual fidelity and realism at extreme-low bandwidth, while maintaining comparable objective fidelity. 
\textit{ResiGLC} holds promise for providing crucial support for robust visual communications in bandwidth-limited and feedback-constrained scenarios.
Future work may focus on multi-modal collaborative compression and concealment, and model compression and quantization for faster inference in practical deployment.

\ifCLASSOPTIONcaptionsoff
  \newpage
\fi

\bibliographystyle{IEEEtran}
\bibliography{Ref}

\vfill
\end{document}